\documentclass[useAMS,usenatbib]{mn2e}
\usepackage{amssymb,graphicx,natbib,longtable,hyperref}
\usepackage{breakurl}

\bibpunct{(}{)}{;}{a}{}{,} 
\newcommand\ion[2]{#1$\;${\small\rmfamily\@Roman{#2}}\relax}%
\newcommand*\tablefootmark[1]{%
  \unskip
  \hbox{\@textsuperscript{\normalfont\ignorespaces#1}}%
  \,%
  \ignorespaces
}
\newcommand\tablefoot[1]{\VSpaceBeforeTabBib=1ex%
  \par\vspace{\VSpaceBeforeTabFoot}
  \noindent
  \begin{minipage}{\linewidth}
    {\aa@tablefootnamefont\aa@tablefootname.}~%
    \aa@tablefootfont
    \ignorespaces
    #1%
  \end{minipage}%
}
\newcommand\tablefoottext[2]{%
  \hbox{\@textsuperscript{\normalfont(\ignorespaces#1)}}%
  ~%
  \ignorespaces
  #2\ \ignorespaces%
}

\newcommand{\sxvi} {S\,{\sc xvi}}

\newcommand{\arxviii} {Ar\,{\sc xviii}}
\newcommand{\caxix} {Ca\,{\sc xix}}
\newcommand{\caxx} {Ca\,{\sc xx}}
\newcommand{\fexxv} {Fe\,{\sc xxv}}
\newcommand{\fexxvi} {Fe\,{\sc xxvi}}

\newcommand{\msun}{M$_{\odot}$}

\usepackage[british]{babel}

\title[GRO J1744-28: an intermediate B-field pulsar in a LMXB]{
GRO J1744-28: an intermediate B-field pulsar in a low mass X-ray binary}

\author[A. D'A\`i et al.]{A.~D'A\`i$^{1}$, \thanks{antonino.dai@gmail.com} 
T. Di Salvo$^{2}$,
R. Iaria$^{2}$,
J. A. Garc\'ia$^{3}$,
A. Sanna$^{1}$,
F. Pintore$^{1}$,
A. Riggio$^{1}$,
\and 
L. Burderi$^{1}$, 
E. Bozzo$^{4}$,
T. Dauser$^{5}$
M. Matranga$^{2}$,
C. G. Galiano$^{2}$,
N.R. Robba$^{2}$
\vspace{6pt}\\
$^{1}$ Universit\`a degli Studi di Cagliari, Dipartimento di Fisica, SP Monserrato-Sestu KM 0.7, 09042 Monserrato, Italy\\
$^{2}$ Universit\`a degli Studi di Palermo, Dipartimento di Fisica e Chimica, via Archirafi 36, 90123 Palermo, Italy\\
$^{3}$ Harvard-Smithsonian Center for Astrophysics, 60 Garden St., Cambridge, MA 02138, USA\\
$^{4}$ ISDC Data Centre for Astrophysics, Chemin dEcogia 16, CH-12 90, Versoix, Switzerland\\
$^{5}$ Dr Karl Remeis-Observatory and Erlangen Centre for Astroparticle Physics, Sternwartstr. 7, D-96049 Bamberg, Germany\\
}

\begin{document}


\pagerange{\pageref{firstpage}--\pageref{lastpage}} \pubyear{0000}

\maketitle

\begin{abstract}
  The \textit{bursting  pulsar}, GRO J1744-28, went  again in outburst
  after $\sim$18 years of quiescence  in mid-January 2014.  We studied
  the broad-band, persistent,  X-ray spectrum using X-ray  data from a
  \textit{XMM-Newton} observation, performed almost at the peak of the
  outburst, and from a  close \textit{INTEGRAL} observation, performed
  3 days  later, thus covering  the 1.3--70.0 keV band.   The spectrum
  shows  a  complex  continuum  shape that  cannot  be  modelled  with
  standard  high-mass  X-ray  pulsar  models,  nor  by  two-components
  models.  We observe broadband and peaked residuals from 4 to 15 keV,
  and we propose a  self-consistent interpretation of these residuals,
  assuming they are produced by cyclotron absorption features and by a
  moderately  smeared,  highly   ionized,  reflection  component.   We
  identify the  cyclotron fundamental  at $\sim$\,4.7 keV,  with hints
  for  two possible  harmonics  at $\sim$\,10.4  keV and  $\sim$\,15.8
  keV. The  position of the  cyclotron fundamental allows  an estimate
  for  the  pulsar  magnetic  field  of  (5.27\,$\pm$\,0.06)  $\times$
  10$^{11}$ G,  if the feature  is produced  at its surface.  From the
  dynamical and relativistic smearing of the disk reflected component,
  we obtain  a lower limit  estimate for the truncated  accretion disk
  inner radius, ($\gtrsim$\,100 R$_g$),  and for the inclination angle
  (18$^{\circ}$--48$^{\circ}$).   We also  detect  the  presence of  a
  softer thermal component,  that we associate with  the emission from
  an accretion disk truncated at a distance from the pulsar of 50--115
  R$_g$.  From  these estimates, we derive  the magneto-spheric radius
  for disk  accretion to be  $\sim$\,0.2 times the  classical Alfv\'en
  radius for radial accretion.
\end{abstract}


\section{Introduction}

Mildly-recycled, slowly spinning, accreting  X-ray pulsars in low-mass
X-ray binaries (LMXBs) are uncommon objects, because of the relatively
short time  required to spin up  the neutron star (NS)  to frequencies
$f_{spin}$  $>$\,100 Hz  during  the X-ray  active  phase. Before,  or
simultaneously, with this \textit{recycling} phase, the magnetic field
(B-field)  of the  NS  decays from  values  B $\gtrsim$\,10$^{12}$  G,
typical of  young NS in  high-mass X-ray  binaries (HMXBs), down  to B
$\lesssim$\,10$^{8}$ G,  when the strength  of the field is  no longer
able to significantly drive the accretion flow and coherent pulsations
in the persistent emission are no longer detected.

Our current understanding  of how NS B-fields  are actually dissipated
is not  yet complete  \citep{payne04, zhang06}, so  that the  study of
LMXBs  where measures  of  the  B-field are  possible  are of  extreme
importance. Direct and  indirect methods for the  determination of the
B-field are  possible if pulsations  are detected,  as in the  case of
recycled LMXBs that host fast spinning NSs (the class of the accreting
millisecond X-ray  pulsars, AMXPs)  and for a  smaller group  of LMXBs
where the  process of recycling is  probably caught in its  very early
stage,   and  the   NS   spin   frequency  is   much   below  100   Hz
\citep{patruno12}.  Members of this class  are e.g.  the 1.7 Hz pulsar
in   X1822-371,   where  the   mass   transfer   is  probably   highly
non-conservative  and   at  super-Eddington   rates  \citep{burderi10,
  iaria13}; the 11 Hz pulsar  in IGR J1748-2446 \citep{papitto11}; the
0.13 Hz pulsar in 4U 1626-67  \citep{beri14}; the 0.8 Hz pulsar in the
well-known system Her  X-1 \citep{tananbaum72} and the  2.14 Hz pulsar
in GRO J1744-28 \citep{degenaar14}.

GRO J1744-28 was discovered in  hard X-rays (25--60 keV) by BATSE/CGRO
on  2nd, December  1995  \citep{kouveliotou96} as  a bursting  source.
Bursts were  soon associated to unstable/spasmodic  accretion episodes
\citep[type-II   bursts,][]{kouveliotou96,lewin96}.   \citet{finger96}
discovered coherent pulsations at 2.14  Hz and its Doppler modulation,
leading to  the determination  of the  orbital period  (11.8337 days),
eccentricity      ($e$\,$<$\,1.1\,$\times$\,10$^{-3}$),      projected
semi-major     axis    (2.6324     lt-s),     and    mass     function
(1.3638\,$\times$\,10$^{-4}$ M$_{\odot}$).  The  pulsed profile showed
a nearly sinusoidal  shape, which together with the value  of the mass
function, implies  the inclination  angle, or  the companion  mass, or
both these conditions, to have low values.

We observed to date three major X-ray outbursts from GRO J1744-28: the
first associated with its discovery  lasting from December 1995 to May
1996, where  at the peak outburst  reached over-Eddington luminosities
\citep{jahoda99,   giles96,woods99,woods00};   the   second   outburst
occurred about  one year later,  started at the beginning  of December
1996  and lasted  about  four  months.  A  long  period of  quiescence
followed, interrupted by  the third and most  recent outburst observed
in January 2014.

It was  soon realized,  just after  its discovery,  that the  low spin
period,  the  long orbital  period  and  the  observed flux  and  spin
derivative  are all  factors  that point  to  an intermediate  B-field
pulsar  (10$^{11}$ G  $<$\,B\,$<$ 10$^{12}$  G) in  a low  inclination
system  \citep{daumerie96}.    \citet{cui97}  presented  observational
evidence for a  propeller effect during the decay period  of the first
outburst,  where  pulsations  were  not   detected  below  a  flux  of
2.3\,$\times$\,10$^{-9}$ erg cm$^{-2}$ s$^{-1}$ (2-60 keV), from which
a  rough limit  on  the Alfv\'en  radius could  be  derived, and  thus
leading to  a B-field  estimate of $\sim$  2.4\,$\times$\,10$^{11}$ G,
assuming a distance of  8 kpc.  \citet{rappaport97} presented possible
evolutionary paths for  GRO J1744-28 that could lead  to the formation
of  the  presently  observed  system,  and  starting  from  a  set  of
reasonable  initial  conditions,  gave  rather  tight  limits  on  the
possible companion mass (0.2--0.7  \msun), neutron star mass (1.4--2.0
\msun),  inclination  angle   (7$^{\circ}$--22$^{\circ}$)  and  pulsar
B-field (1.8--7\,$\times$\,10$^{11}$ G).

Search for the  optical/NIR counterpart has been  difficult because of
the  crowded  Galactic  Centre   field  where  GRO  J1744-28  resides.
$Chandra$   and  \textit{XMM-Newton}   detected   a  low-level   (1--3
$\times$\,10$^{33}$ erg  s$^{-1}$) persistent  X-ray activity  in 2002
\citep{wijnands02,daigne02}, allowing  to refine the estimates  on the
source   coordinates.   \citet{degenaar12}  reported   enhanced   flux
(L$_X$\,$\sim$\,1.9\,$\times$\,10$^{34}$  erg  s$^{-1}$) in  September
2008.    \citet{gosling07}  determined   two  near-infrared   possible
candidate  sources within  the  $Chandra$ error  circle.  Most  recent
determinations  \citep{masetti14atel}  indicate   that  the  candidate
source $a$ by \citet{gosling07} is  the likely counterpart. Hence, the
companion star  is an  evolved G/K  III star.   GRO J1744-28  has been
associated to the Galactic Centre  population, setting its distance at
$\sim$ 7.5 kpc \citep{augusteijn97}.

One of the main peculiarity of GRO J1744-28 is the presence of Type-II
bursts    \citep{lewin96}.     \citet{cannizzo97}    identified    the
Lightman-Eardley instability occurring at the magnetospheric radius in
the accretion  disk as the  possible cause of the  bursting behaviour.
\citet{bildsten97} argued  for the possible presence  of type-I bursts
together with the  type-II as observed also for the  Rapid Burster for
accretion   rates   lower  than   6\,$\times$\,10$^{-9}$   M$_{\odot}$
yr$^{-1}$.  The  bursts have  similar structures,  with a  fast rising
(few seconds), a peak emission that  lasts few seconds and a tail that
appears as  a dip with  respect to  the pre-burst average  count rate.
During the bursts the pulsed fraction increases but the phase with the
pre-outburst  pulsations is  not  maintained, leading  to the  unusual
phenomenon of pulsar glitches \citep{stark96}.

The X-ray continuum emission of  GRO J1744-28 during the 1996 outburst
has  been studied  using  the low  energy-resolution $RXTE$/PCA  data.
\citet{giles96}  modelled this  spectrum using  an absorbed  power-law
with  a high-energy  cut-off (E$_{cut}$\,=\,20  keV, E$_{fold}$\,=\,15
keV).   They   also  found  a  pulse-dependent   photon-index  between
1.20\,$\pm$\,0.02   (at  the   peak  of   the  pulsed   emission)  and
1.35\,$\pm$\,0.01  (at  the  pulse  minimum),  and  detected  a  broad
($\sigma  \sim$\,1 keV)  Gaussian emission  line in  the Fe  K$\alpha$
region.  \citet{nishiuchi99}  presented a  spectroscopic investigation
with  ASCA data  for two  different dates  and accretion  rates.  They
found  the  line  to  be  intrinsically  broad  and  they  proposed  a
relativistically  smeared reflection  component  as the  cause of  the
broadening.   The  iron  line  profile was  compatible  with  resonant
emission from mildly  ionized iron for the  first observation, whereas
highly ionized  (6.7--6.97 keV) iron  seemed preferred for  the second
observation.  However,  the  best-fitting value  for  the  inclination
angle, $>$\,50$^{\circ}$, seems unlikely because of the constraints on
the mass function and companion's  spectral type. Similarly, the value
for   the   inner   disk  radius,   10   R$_g$\,$<$\,R$_{in}$\,$<$\,30
R$_g$\footnote{The   gravitational    radius   R$_g$\,=\,GM/c$^2$   is
  $\sim$\,2  km for  a  canonical 1.4  M$_{\odot}$  NS}) appears  also
unlikely because of  the different indications for  an intermediate NS
B-field  \citep{daumerie96, cui97,  rappaport97}, that  would truncate
the disk at larger distances,  so that other broadening mechanisms, or
a more complex profile, have been additionally advocated.

\citet{degenaar14}    reported   from    a   recent    high-resolution
$Chandra$/HETGS  observation  the  detection  of  an  asymmetric  line
profile  from He-like  iron, and  interpreted it  as a  disk reflected
feature, from which an inner  disk radius of 85\,$\pm$\,11 R$_{g}$ and
an  inclination   angle  of  52$^{\circ}\,\pm\,4^{\circ}$   have  been
estimated.  \citet{younes15}  analysed the  broadband spectrum  of GRO
J1744-28 with  $NuSTAR$ and $Chandra$, fitting  the continuum emission
with  a cut-off  power-law of  photon-index $\Gamma  \sim$\,0, cut-off
energy  $\sim$\,7 keV,  and a  soft continuum  thermal component  at a
temperature of  0.55 keV, interpreted  as emission from  the accretion
disk.  They  also reported the  presence of a  significantly broadened
($\sigma$\,=\,3.7   keV)  feature   at  $\sim$\,10   keV.   Moderately
broadened highly-ionized emission  lines at 6.6 keV and  7.01 keV were
also detected, together with emission from neutral iron at 6.4 keV.

\subsection*{The 2014 outburst of GRO J1744-28}

\begin{figure}
\centering
\includegraphics[height=\columnwidth, width=0.8\columnwidth,  angle=-90]{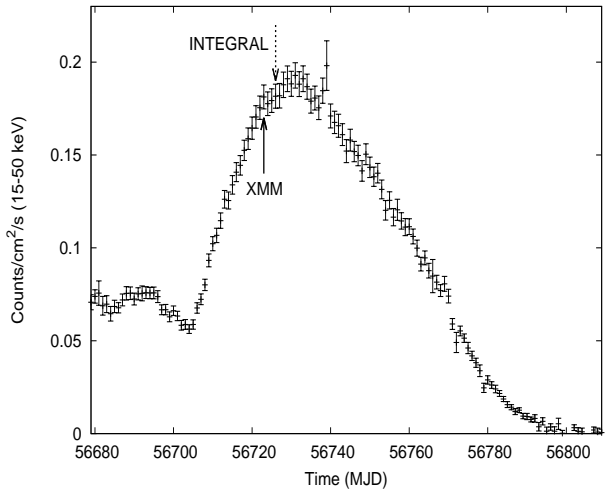}
\caption{$Swift$-BAT \citep{krimm13}  daily light curve in  the 15--50
  keV  range from  22nd January  to 1st  June 2014.   The black  thick
  (dotted)   arrow  marks   the   time   of  the   \textit{XMM-Newton}
  (\textit{INTEGRAL}) observation.}
\label{batlightcurve}
\end{figure}

Enhanced  X-ray   emission  from   the  direction  of   the  transient
outbursting source GRO  J1744-28 was detected from  18th, January 2014
with MAXI  \citep{negoro14atel}.  Pulsations at 2.14  Hz were detected
in  hard  X-rays   with  FERMI/GBM  monitor  on   24th,  January  2014
\citep{finger14atel} and in  soft X-rays with the  $Swift$/XRT on 6th,
February  2014 \citep{dai14atel},  confirming the  association of  the
ongoing outburst with  the bursting pulsar.  The  outburst reached its
peak  in  mid-March,  when  the  source was  accreting  close  to  the
Eddington limit \citep{negoro14b} and from  then it started to decline
linearly  until  23th April  when  the  light  curve  as seen  in  the
$Swift$/BAT monitor (Fig.~\ref{batlightcurve})  clearly showed a drop,
and the  profile became  exponential with  a decay  time of  $\sim$ 11
days.  The outburst was regularly  monitored by $Swift$ and $INTEGRAL$
satellites \citep{masetti14atel}.  A Target of Opportunity observation
with \textit{XMM-Newton}  observed the source from  6th 14:03:12.27 to
7th 13:29:50  March 2014, for a  total exposure time of  83915 s.  The
observation  was  performed few  days  before  the GRO  J1744-28  flux
reached  the  outburst  peak  as   shown  by  the  black  thick  arrow
superimposed  to the  BAT light  curve from  MJD 56679  (22nd, January
2014) to MJD  56809 (1st, June 2014)  in Fig.~\ref{batlightcurve}. For
this work we also exploited data from an \textit{INTEGRAL} observation
(dotted arrow in Fig.  \ref{batlightcurve}) performed three days later
the \textit{XMM-Newton} pointing.

%
%
\section{Observation and data reduction}
We  used the  Science Analysis  System  (SAS) version  13.5.0 and  the
calibration files (CCF) updated at  March 2014 for the data extraction
and analysis  of the  \textit{XMM-Newton} data;  we used  the software
package HeaSOFT version 6.15.1 for the timing analysis and for general
purposes data reduction, and Xspec v.12.8.1 for spectral fitting.

The EPIC/pn  operated in  Timing Mode with  the optical  thick filter,
while the Reflection Grating  Spectrometer instruments (RGS1 and RGS2)
in Spectroscopy Mode.   The EPIC/MOS instruments were  switched off to
allow the highest possible telemetry for the EPIC/pn.

The  RGS  data  were  extracted using  the  \texttt{rgsproc}  pipeline
task. The  total exposure for  the RGS1  (RGS2) instrument is  83715 s
(83750), and  the first-order  net light curves  has an  average count
rate  per second  (cps) of  0.283\,$\pm$\,0.002 (0.268\,$\pm$\,0.002).
Second-order count  rates, for RGS1  and RGS2, are  $\sim$\,30\,\% and
$\sim$\,70\,\% the  first-order rates,  respectively.  The  bursts are
not clearly  resolved in the RGS  light curves, because the  source is
strongly absorbed at lower energies and  the RGS1 and RGS2 frame times
are 4.59  s and 9.05  s, respectively, that  are long compared  to the
rapid burst evolution.

\begin{figure}
\centering
\includegraphics[height=\columnwidth, width=0.8\columnwidth,  angle=-90]{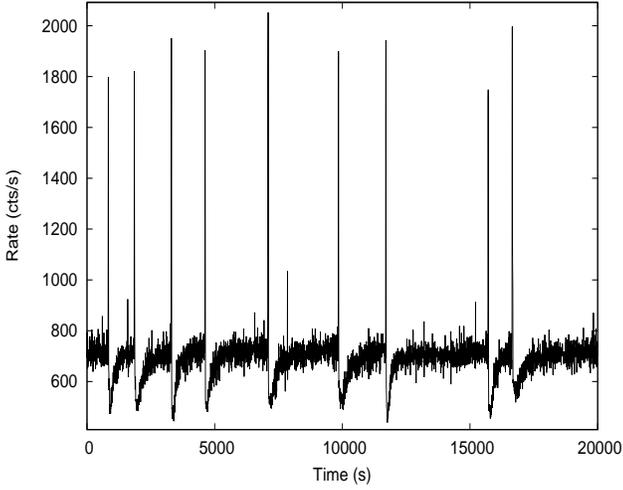}
\caption{EPIC/pn light curve for the first 20 ks
of the observation. Bin time is 5 s.}
\label{pn_lightcurve}
\end{figure}

The  EPIC/pn  event  file  was  processed  using  the  \texttt{epproc}
pipeline   processing   task    with   the   \texttt{runepreject=yes},
\texttt{withxrlcorrection=yes},       \texttt{runepfast=no}.       and
\texttt{withrdpha=yes}, as  suggested by  the most  recent calibration
status
report\footnote{\url{http://xmm2.esac.esa.int/docs/documents/CAL-TN-0018.pdf}}
\citep[see also][]{pintore14}.  The average count rate,  corrected for
telemetry   gaps  (\texttt{epiclccorr}   tool),   during  the   entire
observation over all  the EPIC/pn CCD was 714 cps.   The EPIC/pn light
curve shows 43  bursts during the entire observation with  a mean rate
of 0.53  burst hr$^{-1}$. Fig.~\ref{pn_lightcurve} shows  the first 20
ks  of the  EPIC/pn observation,  where nine  bursts are  present. The
bursts profiles can be clearly  resolved, they have similar structures
with  an average  rising  time  and typical  peak  duration  of a  few
seconds.  During  these peaks the  observed rate becomes so  high that
telemetry saturation and strong pile-up prevent a reliable analysis of
the data.
 
The  burst  decay  is  well  described by  an  exponential  tail  with
decay-times in  the range 2--10  s.  The  count rate after  each burst
drops about 10\% with respect to the pre-burst level and returns to it
slowly  after  few hundred  seconds.   Such  behaviour and  the  burst
characteristics are very  close to the typical  phenomenology shown by
the source in the two past outbursts \citep{giles96}.

\begin{figure}
\centering
\includegraphics[height=\columnwidth, width=0.8\columnwidth,  angle=-90]{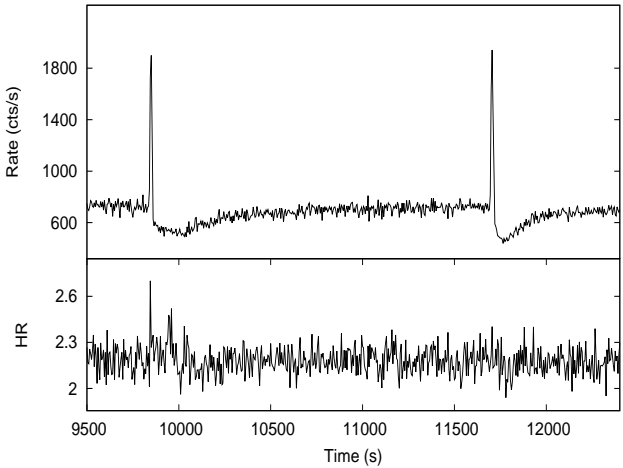}
\caption{Two  consecutive  bursts  and relative  hardness  ratio  (HR,
  4.0--10.0/0.5--4.0  keV).  Selected  extraction region  excludes the
  RAWX\,=\,[36:38] columns to avoid photon pile-up. Bin time is 5 s.}
\label{pn_lightcurve2}
\end{figure}

In Fig.~\ref{pn_lightcurve2} we  show a snapshot of  the EPIC/pn light
curve encompassing two typical bursts  (upper panel), and the hardness
ratio (HR)  evolution, where the soft  color is defined in  the energy
range 0.5--4  keV and  the hard  color in the  4--10 keV  range (lower
panel).   The  light  curve  is  corrected  for  telemetry  saturation
(\texttt{epicclcorr}  task), and  it  is filtered  removing the  three
hottest RAWX columns to limit pile-up effects at the burst peak.

We focused on the persistent  emission of the pulsar by time-filtering
the  EPIC/pn event  list  excluding appropriate  intervals around  the
bursts. The excluded  windows are all of 400 s  length, starting a few
seconds before  the rising  trail of the  bursts. The  final exposure,
corrected  also  for  telemetry  drops,  of  the  persistent  emission
resulted of 64130 s.

\subsection{Timing analysis} \label{timinganalysis}
We barycentred the EPIC/pn, burst-filtered, event file with respect to
the Barycentric  Dynamical Time (TDB) using  the \texttt{barycen} tool
with R.A. and DEC  coordinates 266.137792 and -28.740803, respectively
\citep{gosling07}.  We used  all the events that  passed the filtering
condition  \texttt{FLAG==0}.   A  power  spectrum  of  the  persistent
emission  clearly reveals  the  pulsation  at 2.14  Hz  and the  first
harmonic  at 4.28  Hz.  To  correct the  pulse arrival  times for  the
orbital motion of the pulsar, we used the following orbital parameters
(GBM                                                            pulsar
project\footnote{\url{http://gammaray.msfc.nasa.gov/gbm/science/pulsars/lightcurves/groj1744.html}}):
\textit{P$_{orb}$}\,=\,11.836397 d,  \textit{a~sin(i)}\,=\,2.637 lt-s,
and \textit{T$_{\pi/2}$}\,=\,56695.6988  MJD.  We then used  a folding
search ($efsearch$ in XRONOS) to find the averaged pulse period during
the  persistent  emission.  The  $\chi^2$  highest  peak is  found  in
correspondence with a period of  0.4670450(1) s, or 2.14112130(46) Hz,
where  the uncertainty  corresponds  to the  step  period search  time
(10$^{-7}$ s).

\begin{figure}
\centering
\includegraphics[height=\columnwidth, width=0.8\columnwidth,  angle=-90]{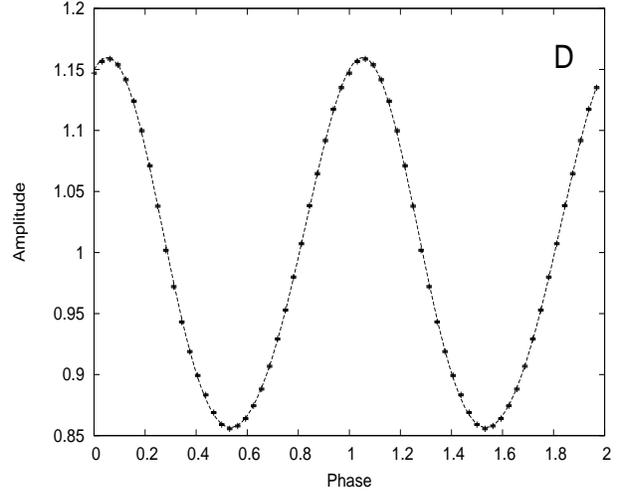}
\includegraphics[height=\columnwidth, width=0.8\columnwidth,  angle=-90]{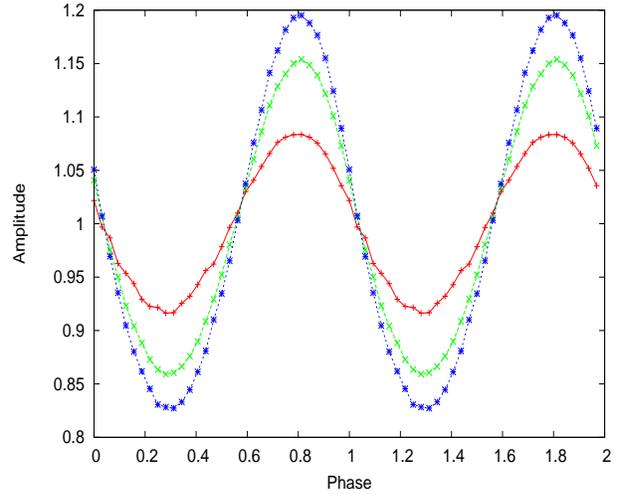}
\caption{   Top  panel:   energy-averaged  pulse-folded   profile  (32
  channels) and best-fitting model using two sinusoids.  Bottom panel:
  pulse-folded profile for the EPIC/pn  data in energy bands: 0.5--3.0
  keV (red data),  3.0--6.0 keV (green data), and  6.0--12-0 keV (blue
  data).}
\label{fig:foldedprofile}
\end{figure}

To check possible shifts in the spin frequency during the observation,
we folded  at this period  the persistent  event file using  a folding
time  window  of 10$^4$  pulse  periods  and  the  start time  of  the
observation as  epoch of reference.   We then plotted the  phases with
respect to this  folding period as a function of  the time. The phases
showed random scattering, and no other correction was found necessary.
Folding the entire persistent emission  at this period resulted in the
folded profile of Fig.~\ref{fig:foldedprofile}. The pulse shape can be
well fitted as  the sum of two sinusoids. Using  the standard notation
for the pulse amplitude $A$,

\begin{equation} \label{eq:amplitude}
A = \frac{I_{max} - I_{min}}{I_{max} + I_{min}} 
\end{equation}
we found the  fundamental has an amplitude of 15.1\%,  while the first
harmonic  amplitude  is  0.91\%,   corresponding  to  $\sim$\,6\%  the
fundamental. The  folded profile is  strongly energy dependent  in the
EPIC/pn   band,   as   clearly   shown  in   the   bottom   panel   of
Fig.~\ref{fig:foldedprofile}, where the folded profile is obtained for
three energy band: 0.5--3.0 keV, 3.0--6.0 keV and 6.0--12.0 keV.

\subsection{The amplitude fraction} \label{sect:amplitude}
We studied  the amplitude  fraction as  a function  of energy  for the
EPIC/pn  band.  We  exploited  all  the selected  events  in the  RAWX
columns,  although  we  were  aware   that  pile-up  may  distort  the
energy-dependent  amplitude measures;  however we  estimated that  the
relative  error  induced  by  some   fraction  of  pile-up  events  is
significantly lower than the statistical uncertainties introduced when
removing the hottest  columns.  We selected the pulse  profiles in the
pulse-invariant (PI) channel  as follows: from 0.5 to 2  keV by a step
of 0.5 keV, from 2 keV to 8 keV  by a step of 0.25 keV, from 8 to 10.5
keV by a  step of 0.5 keV, and  from 10.5 keV to 12 keV  into a single
profile.

\begin{figure}
\centering
\includegraphics[height=\columnwidth, width=0.8\columnwidth,  angle=-90]{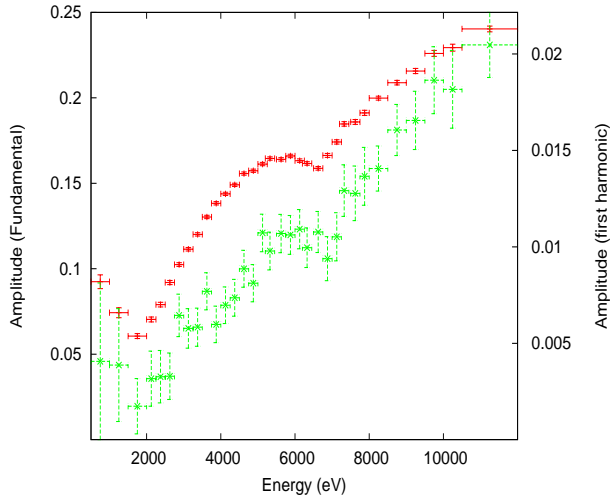}
\caption{Amplitude  fraction  for  the  fundamental  (red  data,  left
  y-axis) and first harmonic (green  data, right y-axis) as a function
  of the energy.}
\label{amplitude}
\end{figure}

We folded  the persistent  event file using  our best  period estimate
(0.4670450  s)  and  fitted  the   resulting  pulse  shape  using  two
sinusoids.   In Fig.~\ref{amplitude}  we  show the  dependence of  the
amplitude of the  fundamental and first harmonic as a  function of the
energy.

\begin{figure}
\centering
\includegraphics[height=\columnwidth, width=0.8\columnwidth,  angle=-90]{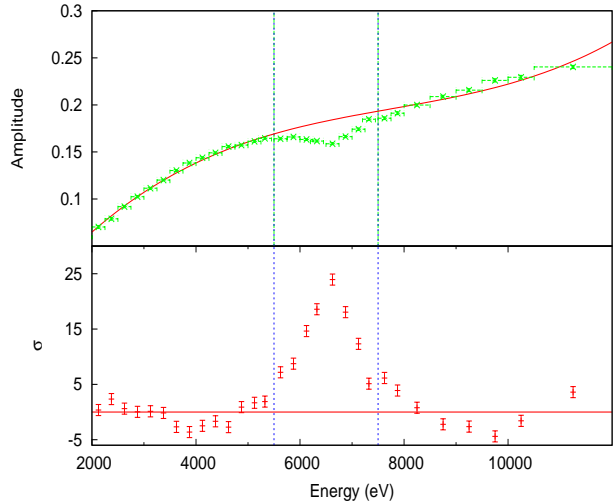}
\caption{Upper panel: fundamental amplitude as a function of energy in
  the 2.0--12.0  keV range, super-imposed with  the best-fitting trend
  derived by data  in the same range, excluding  the interval 5.5--7.5
  keV.  Lower panel: residuals in units  of sigma for the 5.5--7.5 keV
  range  for  the best-fitting  model  consisting  of a  third  degree
  polynomial  derived by  fitting the  data  in the  2.0--5.5 keV  and
  7.5--12 keV range.}
\label{amplitudefit}
\end{figure}

It is evident that above 2 keV the pulse amplitude strongly correlates
with energy,  with the exception of  the energy range between  6 and 7
keV  where a  remarkable  drop  in amplitude  is  present.  To  better
characterize the shape  of this drop we used  a phenomenological model
to describe the  amplitude dependence in the 2.0--5.5  and 7.5--12 keV
range. We  described this  trend using a  polynomial fit,  attaining a
$\chi^2  \sim$  1   with  a  third  order  polynomial.    We  show  in
Fig.~\ref{amplitudefit}, the  residuals of this  fit when the  data of
the 5.5--7.5 keV region are then  noticed.  The shape of the residuals
is  consistent  with  a  Gaussian  profile,  that  it  is  centred  at
$\sim$\,6.5  keV  and has  a  sigma  of  $\sim$\,0.8 keV.   This  drop
indicates the presence of a spectral component that is not pulsed, and
we shall show  that its shape is  compatible with the shape  of a disk
reflection component  as derived  from the  analysis of  the broadband
spectrum (see Sect. \ref{sect:pulsedfraction}).

\section{Spectral analysis}

\subsection{RGS data analysis}

RGS count  rates are  always below the  threshold for  photon pile-up.
Because of  the higher  spectral resolution of  the RGS  detector with
respect to the  EPIC/pn and the low-energy coverage in  a region where
the effect of  ISM absorption is the highest and  the EPIC/pn spectrum
is affected  by larger systematics,  we first performed  a preliminary
spectral analysis in this restricted  energy band.  We extracted first
and second order RGS spectra of the persistent emission using the same
good time intervals obtained from  the EPIC/pn light curve.  We looked
at the background contribution to choose the noticed energy ranges for
spectral  fitting,  keeping  channels  above  $\sim$  3\,$\sigma$  the
background  value. Using  a  simple model  consisting  of an  absorbed
power-law  and  a constant  of  normalization  free  to vary  for  the
different datasets,  we found this inter-calibration  factor among the
four spectra variable at the 7\% level.

We verified the  consistency of the RGS1/RGS2 spectra  and generated a
combined spectrum using the SAS \texttt{rgscombine} tool for the first
(RGSo1) and second (RGSo2) order  spectra.  After having verified that
the  persistent spectrum  is,  within  the statistical  uncertainties,
consistent  with the  total  time-averaged spectrum,  to increase  the
statistical  weight  of the  RGS  data,  we  used this  time  averaged
spectrum  for  the  rest  of   our  analysis.   We  found  no  evident
emission/absorption  narrow feature,  so that  we chose  to group  the
spectra to a  minimum of 100 counts/channel.  Both  spectra (first and
second  orders summed)  are used  in the  1.3--2.0 keV  range, because
below 1.3 keV the spectrum becomes background-dominated.

\subsection{EPIC/pn data analysis: pile-up and background issues}
Because of the high count rate  recorded by the EPIC/pn instrument, we
checked the pile-up level of the persistent time-averaged emission. In
the  EPIC/pn Timing  mode spatial  information is  collapsed into  one
dimension  (RAWX coordinate),  while  spectral  extraction is  usually
performed in a  2-dimensional array, using the  RAWY coordinate, that,
however, is a temporal coordinate.

The point-spread function in the RAWX-RAWY image peaks for the RAWX=37
column and we created four filtered spectra selecting the RAWX=[28:46]
region,  and  successively  removing  one,  three,  and  five  central
columns,   accompanied    always   by   the    filtering   expressions
\texttt{PATTERN$\leq$4} and \texttt{(FLAG==0)}.   We obtained response
matrices using the \texttt{rmfgen} tool  and the ancillary files using
the        \texttt{arfgen}       tool,        following       standard
pipelines\footnote{\url{http://xmm.esac.esa.int/external/xmm_user_support/documentation/sas_usg/USG/epicpileuptiming.html}}.

\begin{figure}
\centering
\includegraphics[height=\columnwidth, width=0.8\columnwidth,  angle=-90]{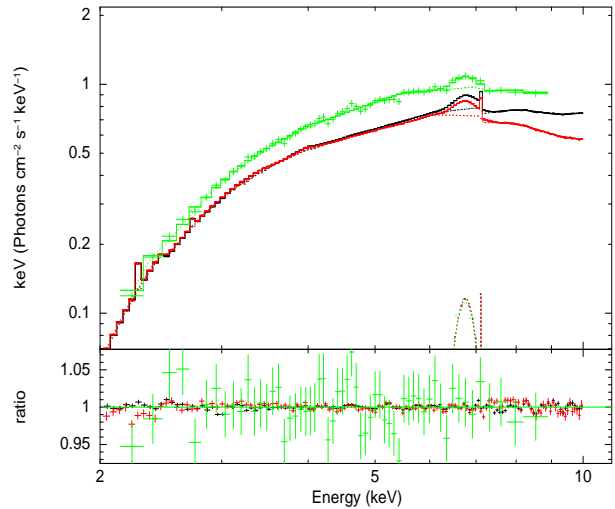}
\caption{Data, unfolded  spectra and data/model ratio  for the EPIC/pn
  spectrum extracted  from the  whole PSF  (no columns  removed, black
  data), for the EPIC/pn spectrum extracted  from the wings of the PSF
  (three columns removed, red data),  and for the $Swift$/XRT spectrum
  from a simultaneous observation (green data).}
\label{pnswift}
\end{figure}

For each  filtered event list,  we applied the  \texttt{epatplot} tool
and we  noted that the exclusion  of three central columns  provided a
spectrum  with  an  observed-to-model   singles  and  doubles  pattern
fraction                            close                           to
unity\footnote{\url{http://xmm.esac.esa.int/sas/current/documentation/threads/epatplot.shtml}}.
However, we  noted that the choice  of excising the PSF  caused severe
effects on  the continuum spectral  determination.  In fact,  when the
spectra  were comparatively  examined,  it resulted  evident that  the
removal of central columns softened  the spectrum beyond what could be
expected  by the  pile-up  correction.   We probed  this  by means  of
spectral analysis. In Fig.\ref{pnswift}, we show the spectra extracted
from    the    EPIC/pn    using   all    the    RAWX[28:46]    columns
(\textit{full~spectrum}),  and  removing  the three  hottest  columns,
together with  a $\sim$ 1 ks  $Swift$/XRT spectrum (WT mode)  taken on
March  7th (OBS.ID  00030898022) in  a simultaneous  time window  with
\textit{XMM-Newton}. We note here that the averaged count rate for the
$Swift$/XRT spectrum  was $\sim$\,70 cps, which  is considerably below
the threshold limit ($\sim$\,100 cps)  for the WT mode photon pile-up,
and the spectrum can be considered a bona fide benchmark.
 
The excised spectrum  clearly displays a strong flux drop  above 7 keV
that is not observed in the  $Swift$/XRT spectrum.  The color ratio of
the  2.0-6.0/6.0-10.0  keV absorbed  flux  is  0.61 for  the  spectrum
without  excision,  and  0.69  for the  spectrum  with  three  columns
removed, while  for the  $Swift$/XRT spectrum we  measured a  value of
0.63, that is  much more consistent with the spectrum  with no removed
column.   We argue  that,  although  some level  of  pile-up could  be
present,  the  choice  of  excising the  PSF  artificially  introduces
spectral distortions,  possibly related  to the reconstruction  of the
energy-dependent  ancillary response,  that  becomes more  significant
than the eventual pile-up correction.   We choose therefore to use the
whole PSF  extraction region for  spectral analysis, but we  will also
show the best-fitting results  for the three-columns excised spectrum,
to have a rough estimate of  the sensitivity due to these instrumental
issues on the  determination of the physical  parameters.  Because the
EPIC/pn   \textit{full~spectrum}  contains   $\sim$\,5$\times$10$^{7}$
counts,  statistical errors  become possibly  dominated by  systematic
errors of  the response matrix  and, to  account for this,  and obtain
final $\chi^2$ close to 1, we  added a 0.5\,\% systematic error to the
EPIC/pn channels \citep[see also][]{hiemstra11}.

It  is suggested  to extract  the background  spectrum from  the first
columns  (e.g.   selecting  RAWX  columns  between 2  and  6)  of  the
RAWX-RAWY image,  however the  flux of the  source is  severe, causing
considerable  spreading  of  the  point  spread  function  leading  to
contamination of  source photons  even in  these first  columns. Light
curves extracted from this region and  from central columns do in fact
show a very  significant positive correlation (p-value\,$<$\,2.2e-16).
To obtain  a measure of  the expected level  of background, we  used a
2001 \textit{XMM-Newton} observation covering the field of view of GRO
J1744-28  (Obs.ID  0112971201)  of  $\sim$ 3\,ks  performed  with  the
EPIC/pn in Imaging Mode (Full-Frame).   In this observation the source
is  quiescent and  no other  point source  is detected  in a  5\arcmin
region around the source coordinates.   We extracted a spectrum around
the coordinates  of the source  using a circular region  of 2.4\arcmin
radius, thus covering the same geometric area of our selection.

We fitted  this spectrum  using a phenomenological  model of  a broken
power-law, obtaining  a satisfactory  description of  the data  with a
reduced  $\chi^2$\,=\,1.1 for  112  d.o.f. We  used this  best-fitting
model to generate a faked  background spectrum using the same response
matrix,  ancillary  response,  and  time  integration  of  the  source
spectrum.  We thus verified  that the expected background contribution
compared  with the  observed  source flux  is very  small  in all  the
EPIC/pn range, contributing $\sim$\,1\%  below 2 keV and $\sim$\,0.2\%
above.   The  background spectrum  extracted  from  the first  columns
contributes $\sim$\,2\% above  2 keV, so although  some variability in
the  region around  the  source can  be expected,  we  argue that  the
background subtraction  may affect the source  flux determination with
this  level  of  uncertainty.  We finally  verified  that  setting  no
background for the EPIC/pn spectrum  had no significant impact for the
determination of the spectral parameters reported in the analysis.

Finally,  we re-binned  the  spectrum in  order to  have  at least  25
counts/channel  and   set  the   oversampling  at   3  channels/energy
resolution with the \texttt{specgroup} tool. We quoted spectral errors
at $\Delta \chi^2$\,=\,2.7, if not stated otherwise.

\subsection{Spectral Analysis of the Persistent Phase-Averaged Emission} \label{sect:nthcomp}

We  first  studied  the  persistent,  pulse-averaged,  spectrum.   For
spectral analysis, we  used the 1.3--2.0 keV range of  the RGS spectra
and the  2.0--9.5 keV range for  the EPIC/pn spectrum. We  first noted
that the EPIC/pn data below 2 keV  do not well match the RGS spectrum,
and  this mismatch  can be  modelled  by an  non-physical soft  excess
present in  the EPIC/pn data  that is not  required in the  RGS.  Such
mismatch has been  reported in many other Timing  Mode observations of
highly   absorbed  (N$_{\rm{H}}$\,$>$\,10$^{22}$   cm$^{-2}$)  sources
\citep[see e.g.][]{dai10}, and it has  been investigated in one of the
most  recent  calibration  technical   reports  of  the  EPIC/pn  team
(CAL-TN-0083). Although we carefully applied the suggested corrections
to the  EPIC/pn processing  pipeline, still  the residuals  do clearly
point to the presence of this instrumental artefact, probably enhanced
by the high statistics of  the EPIC/pn spectrum.  Because the spectrum
below    2    keV    is    well    covered    by    the    RGS    data
(Fig.~\ref{fig:residuals}),  that   do  not  suffer  from   any  known
instrumental  issue,  we  chose  to   discard  EPIC/pn  data  below  2
keV. Because the  source is highly absorbed, we found  that the choice
of the abundance table is  quite relevant for determining the absolute
value of the equivalent hydrogen column and the slope of the continuum
emission as well, given the high correlation between these parameters.
We used the \texttt{tbabs} model and tried all the different abundance
tables                          available                          for
Xspec\footnote{\url{https://heasarc.gsfc.nasa.gov/xanadu/xspec/manual/XSabund.html}},
setting  the  cross-section table  to  that  of \citet{verner96}.   We
derived first some estimates using only  the RGS datasets and a simple
power-law continuum,  obtaining values  for equivalent  column between
(6.24\,$\pm$\,0.07)\,$\times$\,10$^{22}$   cm$^{-2}$,   adopting   the
\citet{anders89}  table,  to  (9.68\,$\pm$\,0.40)\,$\times$\,10$^{22}$
cm$^{-2}$,  adopting the  \citet{wilms00}  abundance  table.  We  then
obtained the corresponding best-fitting  values using only the EPIC/pn
band and finally chose to adopt  the abundance table that provided the
best match  between the RGS and  EPIC/pn energy bands for  the rest of
our analysis (i.e. abundance \texttt{aneb}, from \citet{anders82}).

Because the spectrum  of the source extends well above  10 keV and the
cut-off energy  of the spectrum  is close to  the edge of  the EPIC/pn
band, it  is extremely useful,  to better constrain the  overall X-ray
continuum  shape,  to  complement the  \textit{XMM-Newton}  data  with
observations from other satellites covering the hard X-ray emission of
the   source.     To   this    aim   we   used    publicly   available
\textit{INTEGRAL}/ISGRI      (20.0--70.0      keV      range)      and
\textit{INTEGRAL}/JEMX (we  used both  instruments JEMX1 and  JEMX2 in
the 7.0--18 keV  range) of the $INTEGRAL$ observation  from March 10th
(Revolution  \#1392), three  days after  the $XMM-Newton$  observation
(see   Fig.~\ref{batlightcurve},  where   the   $Swift$/BAT  rate   is
0.181\,$\pm$\,0.006 and 0.182\,$\pm$\,0.006 cts cm$^{-2}$ s$^{-1}$ for
the   dates  of   the   $XMM-Newton$   and  $INTEGRAL$   observations,
respectively).  We  used a calibration  constant to take  into account
flux    variations    occurred     between    the    $INTEGRAL$    and
\textit{XMM-Newton} observations, assuming that the spectral shape has
not significantly varied.

We  started selecting  the  possible continua  based  on the  physical
characteristics of GRO J1744-28, that  is quite likely an intermediate
B-field source,  accreting via  Roche-lobe overflow, with  presence of
strong pulsed emission  (15\% in the average 1.0--10  keV range).  The
pulsed fraction emission is  intermediate between typical values found
in HMXB  pulsars and the  ones found in  AMXPs, and this  supports the
hypothesis that  the NS magnetic field  is not so high  to efficiently
capture matter  from large distance  in the disk. We  expect therefore
that part  of accreting matter  is channelled and up-scattered  in the
accretion column, and that a (significant?) part of the X-ray emission
may  be produced  either at  the magnetospheric  boundary or  directly
linked with emission from the accretion disk (if the disk is truncated
relatively not far  away) or from reflection by the  accretion disc of
the primary radiation.

However, if  the magnetospheric radius R$_{mag}$  is $\gg$\,100 R$_g$,
we expect a  spectrum closer to that shown  from HMXBs.  Consequently,
following the  phenomenological descriptions  mostly adopted  to model
the spectra of accreting HMXBs, we tried to fit the data with a set of
commonly used  spectral models for accreting  X-ray pulsars \citep[see
  model   details   in][]{coburn02}:   a   cut-off   power-law   model
(\texttt{cutoffpl}),  a power-law  with  a  high-energy cut-off  model
(\texttt{highecut}),   a   power-law   with  a   Fermi-Dirac   cut-off
(\texttt{fdco}),  and  a  negative-positive  cut-off  power-law  model
(\texttt{npex}).

Alternatively,  assuming  the magnetic  field  is  relatively low  and
unable  to  stop  the  accretion flow  at  large  distance  (R$_{mag}$
$\lesssim$\,100  R$_g$)  from   the  NS,  we  fitted   the  data  with
two-component models  as commonly  done for low-inclination  angle and
high-$\dot{M}$ accreting LMXBs.

We  used for  the  hard X-ray  emission from  the  accretion column  a
thermal       Comptonization      model,       \citep[\texttt{nthcomp}
  model][]{zycki99},  and  for  the   soft  X-ray  emission  either  a
black-body    component   or    a    multi-coloured   disk    emission
(\texttt{ezdiskbb} component in Xspec, model \texttt{disk+nthcomp}).

\begin{figure}
\centering
\includegraphics[height=\columnwidth, width=0.8\columnwidth,  angle=-90]{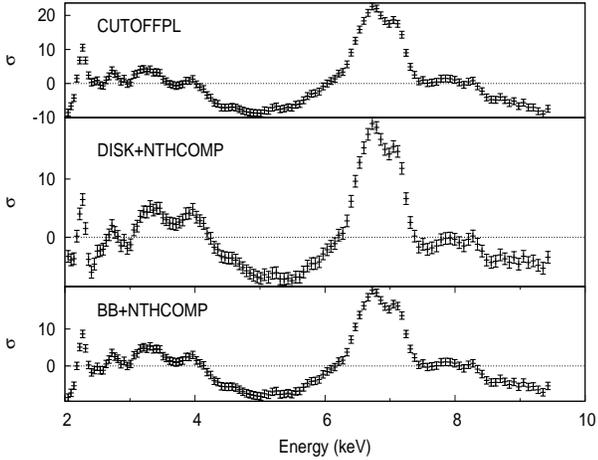}
\caption{EPIC/pn  2--9.5   keV  residuals  for   three  representative
  best-fitting continua.}
\label{bband_res}
\end{figure}

\begin{figure*}
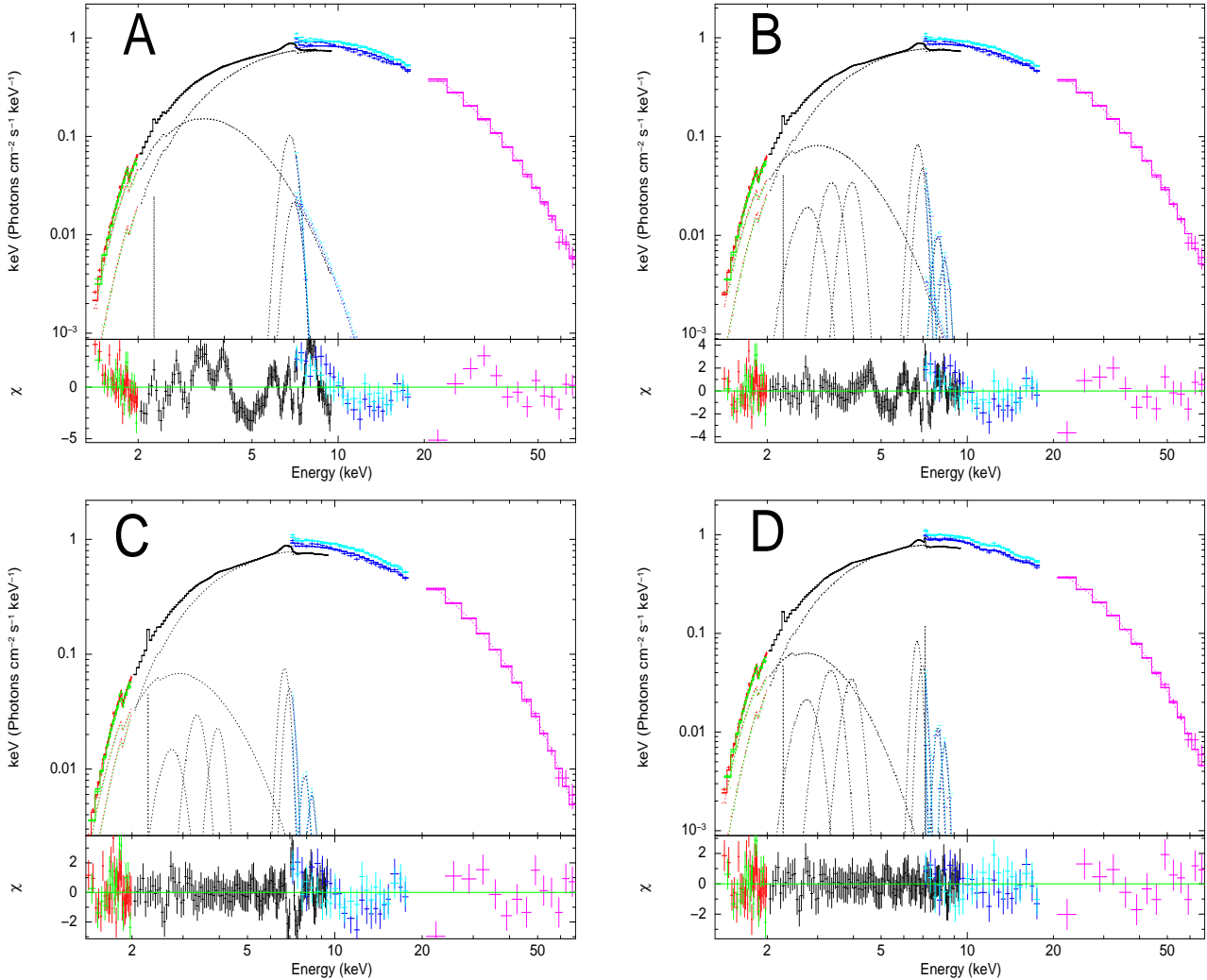

\centering
\begin{tabular}{cc}
\includegraphics[height=\columnwidth, width=0.8\columnwidth,  angle=-90]{fig10.ps} &
\includegraphics[height=\columnwidth, width=0.8\columnwidth,  angle=-90]{fig11.ps} \\
\includegraphics[height=\columnwidth, width=0.8\columnwidth,  angle=-90]{fig12.ps} &
\includegraphics[height=\columnwidth, width=0.8\columnwidth,  angle=-90]{fig13.ps} \\
\end{tabular}
\caption{Data,   best-fit    unfolded   models   ($E    \times   f(E)$
  representation)  and residuals  in  units of  $\sigma$  for a  model
  composed of an accretion disk component and a thermal Comptonization
  model.   RGSo1   data  in   red,  RGSo2   data  in   green,  EPIC/pn
  (\textit{full spectrum})  data in black,  JEMX1 data in  blue, JEMX2
  data  in light  blue, ISGRI  data in  magenta. RGS  data graphically
  re-binned  for  clarity.   Panel   A:  addition  of  two  moderately
  broadened  emission  Gaussian lines  in  the  iron range;  panel  B:
  addition  of  a set  of  highly  ionized  emission lines;  panel  C:
  addition of a \texttt{gabs} at  $\sim$\,5 keV.  panel D: addition of
  two other  \texttt{gabs} harmonics at $\sim$\,10  and $\sim$\,15 keV
  and of a zero-width line at 7.11 keV.}
\label{fig:residuals}
\end{figure*}

We generally found very high reduced $\chi^2$ ($>>$\,10), irrespective
of the continuum choice, with a  common trend in the residuals, driven
by  the  high  signal-to-noise-ratio  of the  EPIC/pn  channels.   The
residuals pattern has  a clear broad peak centred in  the 6.4--7.0 keV
K$\alpha$ iron range, a residuals dip from $\sim$\,4 to $\sim$\,6 keV,
and some local peaks at $\sim$\,2.2 keV, 2.7 keV, 3.3 keV and 4.0 keV.
The residuals  of the  iron peak  strongly pivot the  fit and  cause a
general  softer  continuum and  a  poor  determination of  the  energy
cut-off.  We  show in  Fig.\ref{bband_res}, the EPIC/pn  residuals for
three representative models.

Because  the  residuals  cover  a   large  energy  range,  it  is  not
straightforward to determine their  origin from a single corresponding
local process.  To  simplify the presentation of our  results, we show
and discuss them  in the frame of one  representative continuum model;
we  chose  the \texttt{disk+nthcomp}  model  because  it provided  the
marginal final best $\chi^2$ and  it gives a reasonable representation
of the main physical characteristics  of the spectrum. We repeated the
analysis using  the set of  continua listed  above, and we  found only
small variations  in the spectral  parameters that did not  affect the
overall picture here presented.

We first assumed that the major processes are in emission and tried to
model  the residuals  with a  set  of local  emission Gaussian  lines,
starting by  the highest peak  found in  the iron range.   Because the
residuals indicated a broadened feature,  and a single Gaussian proved
inadequate  to fit  it, we  assumed a  possible blending  of different
transitions and  we checked this  hypothesis using two  Gaussian lines
with fixed energies at 6.7 keV  (He-like iron) and at 6.97 keV (H-like
iron),  tying their  widths, assuming  they are  produced in  the same
photo-ionised plasma.  Both lines are strongly required, with a common
width of $\sim$\,0.35 keV (Fig.\ref{fig:residuals}, panel A), although
the residuals landscape was still far from being satisfactory.

We further examined  the secondary pattern of  local residuals outside
the iron  range. Because  the peaked  emission in  the iron  range had
suggested presence  of fluorescence lines of  highly ionized elements,
we  followed this  scenario and  looked  at the  expected energies  of
H-like and He-like  transitions of the most abundant  elements (Si, S,
Ar, Ca, and Fe).  We added a set of Gaussian lines fixing the energies
at the expected  laboratory frame. To limit the degrees  of freedom of
the  fit  we set  a  common  line  width for  the  Si,  S, Ar  and  Ca
transitions and  another one  for the Fe  transitions. Because  of the
intensity of  the iron lines, we  also took into account  the possible
presence of the K$\beta$ transitions.  We found significant detections
($\geq$ 2 $\sigma$) for the  following transitions: S\,{\sc xvi} (2.62
keV), Ar\,{\sc xviii} (3.32 keV),  Ca\,{\sc xix} (3.90 keV), K$\alpha$
Fe\,{\sc xxv} (6.7 keV), K$\alpha$ Fe\,{\sc xxvi} (6.97 keV), K$\beta$
Fe\,{\sc xxv} (7.88 keV) K$\beta$  Fe\,{\sc xxvi} (8.25 keV).  We also
found a significant narrow line detection at $\sim$\,2.2 keV This line
is found in  correspondence with a strong derivative  of the effective
area at  an instrumental Au M-edge,  and we retain it  of instrumental
origin, so that we  will keep them for the evaluation  of the fits but
we will not longer discuss it.

This  best-fitting model  and the  residuals appeared,  however, still
unsatisfactory, because of the persistent S-shaped curvature between 4
and 6 keV (see Fig.\ref{fig:residuals}, panel B).  A possible physical
process able to impress on the spectrum such features is the cyclotron
resonant  scattering   process  (CRSFs).   CRSFs,  when   besides  the
fundamental, higher harmonics are detected, are present in the spectra
of  young X-ray  pulsars hosted  in  HMXBs, and  are usually  detected
between 20 and 60 keV, whereas their detection in the spectra of X-ray
pulsars  in  LMXBs is  strongly  hampered  because of  lower  B-fields
strengths and/or possible different accretion geometries.

The position of the line, $E_{\rm cycl}$ is related to the strength of
the field by the simple relation:
\begin{equation} \label{eq:cyclotron}
E_{\rm cycl} = 11.6 \times B_{12} / (1 + z) \quad \textrm{keV}
\end{equation} 
where  the B-field  is  expressed  in units  of  10$^{12}$  G and  the
gravitational red-shift  $z$ is a  function of the distance  where the
line is effectively  produced ($z$ $\sim$\,0.3 for  a line originating
very close to the NS surface).  For  the case of GRO J1744-28, where a
possible intermediate  field between 1.8 and  7\,$\times$\,10$^{11}$ G
has been claimed, the presence of a line between 1.8 and 6.2 keV would
be not surprising.

To  model this  feature we  used an  absorption Gaussian  line profile
(\texttt{gabs} component in Xspec) that has three free parameters: the
line energy  position (E$_{1}$),  the width  (W$_{1}$) and  the depth,
D$_{1}$,  related  to  the  optical depth,  $\tau$,  by  the  relation
$\tau$\,=\,0.4$ \times D_{1}/W_{1}$. We  found a strong improvement in
the $\chi^2$ for the addition  of this component (the $\chi^2$/degrees
of  freedom [dof]  passing from  632/524 to  571/521), with  an F-test
probability   of   chance  improvement   of   2\,$\times$\,10$^{-11}$.
Further,  we also  found possible  hints  for the  presence of  higher
harmonics at $\sim$\,10  keV and at $\sim$\,15 keV,  however the depth
and  the position  of these  features showed  strong correlation  with
other  spectral parameters,  resulting in  large relative  errors.  To
limit  these  correlations and  the  implied  large uncertainties,  we
imposed the width  of the first (second) harmonic to  have two (three)
times the value of the  corresponding fundamental width, and with this
constraint,  we  found  a  reduced  $\chi^2$ of  1.052  for  517  dof,
resulting, however,  in a  marginal detection  at $\sim$  3.8 $\sigma$
level.   Finally,  we noted  a  very  narrow  emission line,  with  an
equivalent width  of 9  eV, close  to the energy  of the  neutral iron
edge. The addition  of this feature was found  very significant (final
reduced    $\chi^2$    1.021    for    515    dof,    panel    D    of
Fig.\ref{fig:residuals}).   After  this  last component  addition,  no
other evident feature was apparent in the residual pattern. The values
of  the instrumental  constants for  the different  data-sets resulted
generally reasonable.   There is  a $\sim$  10\% mismatch  between the
EPIC/pn data and the RGS first-order/second-order data, and it is most
likely  due to  uncertainties  in the  reconstruction  of the  EPIC/pn
effective   area   and   to   the  EPIC/pn   background   issues.    A
cross-calibration factor, around 30\%, is obtained between the EPIC/pn
and the $INTEGRAL$/JEMX2 and $INTEGRAL$/ISGRI data, but these data are
not simultaneous  with the  EPIC/pn data as  the $INTEGRAL$  data were
taken few days later the $XMM-Newton$ observation.

The  continuum emission  is characterized  by three  temperatures.  We
associated  the  softest  (kT$_{disk}$\,=\,0.54 keV)  to  the  maximum
temperature reached in a multi-coloured accretion disk, and we derived
from the  normalization of the  component (N$_{ez}$), an  estimate for
the accretion disk inner radius (R$_{ez}$) through the relation:

\begin{equation} \label{eq:disknorm}
N_{ez} = cos(i) \left( \frac{R_{ez}}{D f^2} \right)^2 	
\end{equation}

where $D$  is the source  distance and $f$  is the color  to effective
ratio.    We   assumed   for   this  estimate   a   $D$\,=\,7.5   kpc,
$i$\,=\,20$^{\circ}$  and  $f$\,=\,2,  obtaining a  best-estimate  for
R$_{ez}$ of 27.5 R$_{g}$.

The second temperature (kT$_{bb}$\,=\,1.8 keV) is the soft seed photon
temperature and  it is related  to the radiation field  which provided
the source  of soft  photons, and  it could  correspond to  the plasma
temperature   in  the   post-shock  region,   or  the   \textit{mould}
temperature at  the surface of  the NS \citep{becker07}.   The highest
temperature (kT$_e$\,=\,6.7 keV) is  the thermal electron temperature,
responsible  for  the  Compton   up-scattering  of  the  soft  photons
(neglecting the  contribution from bulk-motion Comptonization)  and it
is associated  with the spectral  cut-off energy.  If we  consider the
RGSo1 flux  as a more  reliable proxy  for estimating the  source flux
(because it is likely much less  affected than the EPIC/pn spectrum by
background and  systematic uncertainties), we derived  a 1.0--10.0 keV
unabsorbed flux  of 1.32\,$\times$\,10$^{-8}$ erg  cm$^{-2}$ s$^{-1}$,
and an extrapolated 0.1--100  keV flux of 3.0\,$\times$\,10$^{-8}$ erg
cm$^{-2}$ s$^{-1}$.  For  a distance of 7.5 kpc,  the isotropic source
luminosity is $\sim$\,2.1\,$\times$\,10$^{38}$  erg s$^{-1}$, slightly
above  the  Eddington  limit  for   a  canonical  1.4  M$_{\odot}$  NS
(L$_{Edd}$ $\sim$  1.8\,$\times$\,10$^{38}$ erg s$^{-1}$).  Within the
energy  band covered  by the  data  (1.3--70 keV),  the relative  flux
contribution of the disk component is  6.2\% of the total X-ray output
from source.

The final  best-fitting model and  residuals are  shown in panel  D of
Fig.~\ref{fig:residuals},  the best-fitting  parameters and  errors of
the continuum emission are shown in Table~\ref{tab:spectral_fits}, and
the best-fitting parameters  and errors for the set  of emission lines
in Table~\ref{tab:lines}.  We determined the uncertainties on the line
energies and widths using the  EPIC/pn data set without any systematic
error, because these parameters are not  expected to be driven by this
type of uncertainty.

\begin{table}
\caption{Spectral fitting results for the model \texttt{disk+nthcomp}. Continuum and 
cyclotron line parameters. Errors given at $\Delta \chi^2\,=\,2.706$ (90\% confidence
for the single interesting parameter).}
\centering
\begin{tabular}{ll|l}
\hline
Parameter (Units) &          \multicolumn{2}{c}{Value} \\ \hline
N$_{	\textrm H}$  (10$^{22}$ cm$^{-2}$)      & 6.3$\pm$0.1    \\
$\Gamma$                                    & 1.96$\pm$0.09  \\  
kT$_e$   (keV)                              & 6.75$\pm$0.26  \\
kT$_{bb}$  (keV)                            & 1.80$\pm$0.08  \\
Norm.$^{a}$                                 & 0.11$_{-0.01}^{+0.04}$  \\
\hline
kT$_{disk}$   (keV)                     &   0.54$_{-0.07}^{+0.11}$  \\ 
Inner disk radius R$_{ez}$ (R$_{g}$)    &   27$_{-7}^{+18}$   \\
\hline
E$_{1}$    (keV)                   & 5.31$_{-0.14}^{+0.09}$  \\
W$_{1}$      (keV)                 & 0.39$_{-0.09}^{+0.18}$   \\
D$_{1}$      ($\times$ 10$^{-2}$)   & 2.9$_{-0.1}^{+0.4}$    \\
\hline
E$_{2}$  (keV)                      & 11.4$_{-0.4}^{+1.0}$ \\
W$_{2}$ (keV)                       & 0.78$^{b}$  \\
D$_{2}$  ($\times$ 10$^{-2}$)        & 22$\pm$10  \\
\hline
E$_{3}$  (keV)                       & 15$_{-1.0}^{+2.0}$ \\
W$_{3}$  (keV)                       & 1.17$^{c}$ \\ 
D$_{3}$ ($\times$ 10$^{-2}$)          & 32$_{-15}^{+25}$      \\
\hline
Constant RGSo1/PN  & 1.13$\pm$0.02  \\
Constant RGSo2/PN  & 1.08$\pm$0.02 \\
Constant JEMX1/PN  & 1.18$\pm$0.02 \\ 
Constant JEMX2/PN  & 1.32$\pm$0.02 \\
Constant ISGRI/PN  & 1.33$\pm$0.10 \\
\hline
$\chi^2_{red}$ (dof)                & 1.00 (515)   \\
\hline \hline
\multicolumn{2}{l}{$^a$ \texttt{nthcomp} normalization, unity at 1 keV for a norm of 1.}\\
\multicolumn{2}{l}{$^b$ Parameter tied to 2\,$\times$\,W$_{1}$.}\\
\multicolumn{2}{l}{$^c$ Parameter tied to 3\,$\times$\,W$_{1}$.}\\
\end{tabular}
\label{tab:spectral_fits}
\end{table}

\begin{table*}
\center
\caption{Emission lines in the spectrum of GRO J1744-28. Errors given at $\Delta \chi^2\,=\,2.706$ (90\% confidence
for the single interesting parameter).}  
\begin{tabular}{lll lll}
\hline \hline
Ion (Transition$^{a}$) & E$_{lab}$& E$_{obs}$     & Width & Flux$^{b}$  & EQW$^{c}$ \\
                       & keV      & keV           & eV    & (10$^{-4}$) &  eV    \\
\hline \hline
K$\alpha$ \sxvi    & 2.62 &  2.65$_{-0.03}^{+0.04}$    & =\caxix     & 160$_{-10}^{+40}$  & 60$_{-10}^{+20}$\\ 
K$\alpha$ \arxviii & 3.32 &  3.28$_{-0.02}^{+0.03}$    & =\caxix     & 150$_{-10}^{+40}$  & 70$\pm$5 \\ 
K$\alpha$ \caxix   & 3.90 &  3.91$\pm$0.02             & 240$\pm$10  & 82$_{-17}^{+10}$   & 44$\pm$6 \\
K$\alpha$ \fexxv   & 6.70 & 6.69$\pm$0.04   & 210$_{-40}^{+20}$ & 67$_{-11}^{+30}$   & 55$\pm$15 \\
K$\alpha$ \fexxvi  & 6.97 & 6.96$\pm$0.07   & =\fexxv           & 40$_{-22}^{+12}$   & 30$\pm$10  \\ 
K$\beta$ \fexxv     & 7.88 & 7.88$^{d}$  & =\fexxv              & 7.9$_{-3.2}^{+4.9}$ & 7.4$\pm$2.5 \\     
K$\beta$ \fexxvi    & 8.25 & 8.25$^{d}$  & =\fexxv              & 5.1$_{-1.3}^{+1.7}$ & 5.1$_{-1.0}^{+2.3}$  \\ 
\hline 
systematic         &      &  2.264$\pm$0.004           & 0$^{d}$     & 60$\pm$6  & 18$\pm$3 \\ 
systematic         &      &  7.11$_{-0.1}^{+0.2}$      & 0$^{d}$     & 12$_{-1}^{+2}$ &  9.4$\pm$2.4   \\
\hline
\hline
\multicolumn{6}{l}{$^a$ Rest-frame energies from \citet{verner96}.}\\
\multicolumn{6}{l}{$^b$ Total area of the Gaussian (absolute value), in units of photons/cm$^{2}$/s.}\\
\multicolumn{6}{l}{$^c$ Line equivalent width.}\\
\multicolumn{6}{l}{$^d$ Frozen parameter.}\\
\end{tabular}
\label{tab:lines}
\end{table*}
\subsection{Modelling the highly-ionized features with a reflection component} \label{sect:xillver} 

We have shown in the previous section as a possible interpretation for
the  large deviations  from  the continuum  shape can  be  due to  the
presence  of  large,  moderately  deep, absorption  features  that  we
interpreted as signs of CRSFs. We have also shown a complex pattern of
local emission  features, moderately broadened, and  compatible with a
set of  highly ionized  transitions from a  photo-ionized environment.
We  investigate  the characteristics  of  the  line emission  spectrum
applying a self-consistent and  physically motivated scenario for this
set  of local  features.  Excluding  a  possible origin  in a  thermal
environment, that  would require unrealistic MeV  temperature, we make
the  hypothesis that  this set  of lines  is due  to reflection  in an
accretion disk  that is still  relatively close  to the NS  to produce
appreciable dynamical and relativistic broadening.

To  this aim,  we slightly  changed the  continuum shape  in order  to
self-consistently incorporate  the reflection component,  adopting the
reflection      model     \texttt{relxill}      \citep[see     details
  in][]{dauser10,garcia14}.  This model  assumes as incident continuum
a  cut-off   power-law  and   it  correctly  takes   into-account  the
angle-dependent  reflection  pattern.   The  reflection  component  is
calculated only down  to values of photon-index, $\Gamma$,  of 1.4 and
cut-off energy of 20 keV; such  values are sensibly different from the
continuum parameters  that we  found, however, because  the reflection
component  is mainly  a contributor  in the  iron range  (fluorescence
lines),  we  retain the  model  still  a reasonable  and  satisfactory
approximation  of  the  true  reflection pattern.   We  left  as  free
parameter the inner disk radius (R$_{in}$\footnote{To be noted that we
  drew another  independent estimate  for the  same quantity  from the
  best-fitting  model  derived  from  the normalization  of  the  disk
  component, that we  labelled R$_{ez}$, Eq.~\ref{eq:disknorm}.}), the
inclination angle  and the  emissivity index ($\epsilon$).   The outer
disk radius  was held fixed  at 1000  R$_{g}$, and the  spin parameter
$a$=0.  The other free parameters of this component are the ionization
parameter (log($\xi$)), the reflection  fraction ($Refl.~fr.$) and the
iron abundance relative to the solar value ($A_{Fe}$).

\begin{figure}
\centering
\includegraphics[height=\columnwidth, width=0.8\columnwidth,  angle=-90]{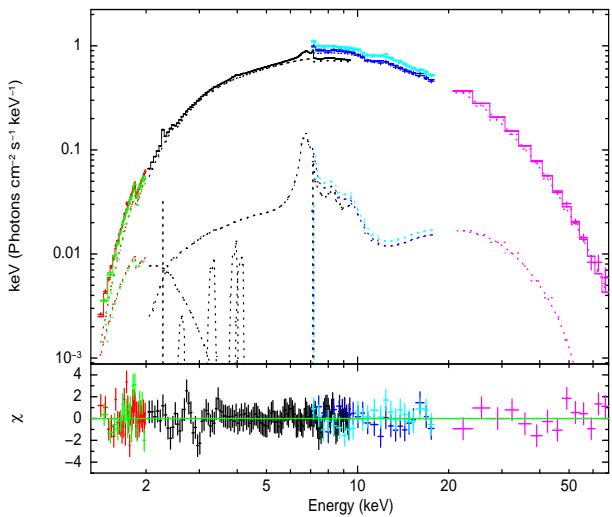}
\caption{Data,  unfolded model with additive components  (upper  panel) and residuals in units of $\sigma$ (lower
  panel)       using       the       \texttt{disk+relxill}       model1
  (Table~\ref{tab:xillverP}) and a set  of ionized emission lines with
  common blurring parameters.}
\label{fig:xillver}
\end{figure}

We tested our constraints on the widths of the CRSFs, but we found the
widths of the first and  second harmonic to be significantly different
from 2\,$\times$\,W$_{1}$  and 3\,$\times$\,W$_{1}$, and, to  obtain a
satisfactory fit, we left the CRSFs  widths free to vary. We note that
the two harmonics become statistically  more necessary with respect to
the continuum  model of  Table \ref{tab:spectral_fits}  (both features
detected  at  $\sim$\,6.8\,$\sigma$,  but  the  single  harmonics  are
detected   at   $\sim$\,3\,$\sigma$   level),   implying   a   certain
model-dependent detection.  We tested the  presence of the softer disk
component, left free to vary energy  and normalization of the 2.22 keV
instrumental  feature.  The  best-fitting  model has  a final  reduced
chi-squared of 1.32  (526 dof), it describes  quite satisfactorily the
residual pattern  in the iron range  although we noted (left  panel of
Fig.\ref{fig:xillver})  that some  residuals  still  persisted at  the
energies  corresponding  to  the  K$\alpha$  Si\,{\sc  xvi},  Ar\,{\sc
  xviii}, Ca\,{\sc xix} and Ca\,{\sc xx} emission lines.  To model the
larger  intensities  suggested by  these  residuals,  we used  blurred
relativistic profiles (\texttt{relconv} model) for the non-iron highly
ionized  emission lines,  setting the  blurring parameters  (inner and
outer disk radius,  inclination angle and emissivity  profile index of
the \texttt{relconv}  component) to be  tied to the parameters  of the
\texttt{relxill}  model.    We  found  that  the   broadened  line  at
$\sim$\,3.9  keV,   identified  with   the  K$\alpha$   Ca\,{\sc  xix}
transition in  Table~\ref{tab:lines}, is  much better modelled  with a
combination of He-  and H-like calcium emission lines.   We found that
the  energies of  the  K$\alpha$ Ar\,{\sc  xviii},  Ca\,{\sc xix}  and
Ca\,{\sc xx} are  consistent with the expected  rest-frame values, but
the position  of the  S\,{\sc xvi}  line is  found at  higher energies
(2.73\,$\pm$\,0.02 keV), however  setting the line energy  at 2.62 keV
still gives  a lower limit on  the normalization higher than  zero. We
calculated the  errors on  the line  normalizations freezing  the line
energies  at  the  laboratory frame  values,  obtaining  normalization
values   significantly    lower   than   the   values    reported   in
Table~\ref{tab:lines} and physically more consistent with the expected
widths implied  by the  blurring parameters  of the  fluorescence iron
transitions (Table~\ref{tab:xillverP}).

\begin{figure*}
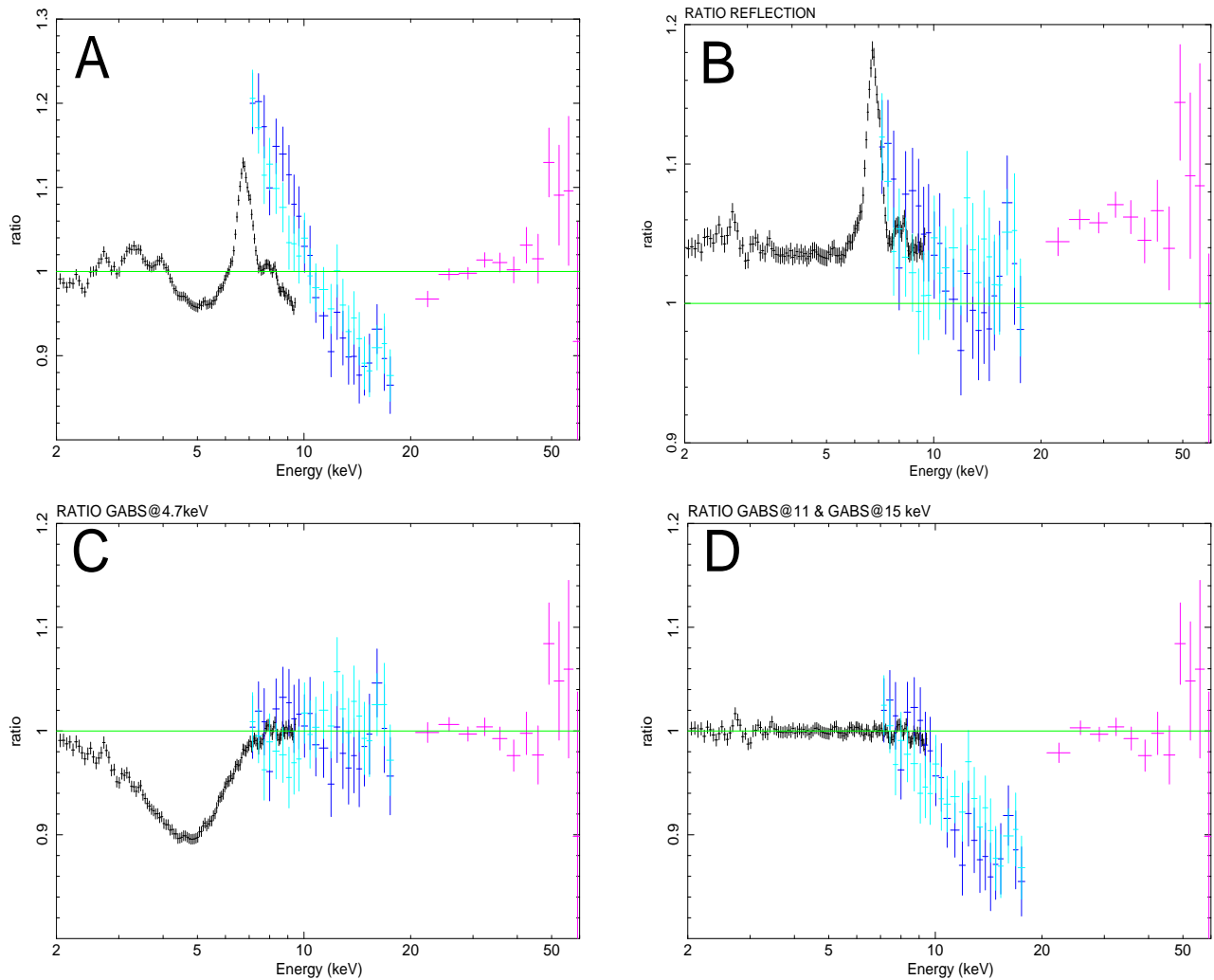

\centering
\begin{tabular}{cc}
\includegraphics[height=\columnwidth, width=0.8\columnwidth,  angle=-90]{fig15.ps} & 
\includegraphics[height=\columnwidth, width=0.8\columnwidth,  angle=-90]{fig16.ps} \\
\includegraphics[height=\columnwidth, width=0.8\columnwidth,  angle=-90]{fig17.ps} &
\includegraphics[height=\columnwidth, width=0.8\columnwidth,  angle=-90]{fig18.ps} \\
\end{tabular}
\caption{2-60  keV data/model  ratios:  for  a best-fitting  continuum
  without reflection and \texttt{gabs} (panel A), for the best-fitting
  continuum  of Table~\ref{tab:xillverP}  with $Refl.\,fr.$  set to  0
  (panel     B);     for     the     best-fitting     continuum     of
  Table~\ref{tab:xillverP} with D$_{1}$ set to 0; for the best-fitting
  continuum of  Table~\ref{tab:xillverP} with D$_{2}$ and  D$_{3}$ set
  to 0.}
\label{fig:ratios}
\end{figure*}

Finally,  we added  a  zero-width line  at  $\sim$\,7.11 keV,  finding
best-fitting  position and  normalization consistent  with the  values
reported  in Table~\ref{tab:lines},  and  obtaining  a final  best-fit
reduced  $\chi^2$ of  1.025 (519  dof) and  a satisfactory  pattern of
residuals     as      shown     in      Fig.~\ref{fig:xillver}.     In
Fig.~\ref{fig:ratios}, we  show the  broadband data/model ratio  for a
best-fit continuum without emission/absorption features (panel A); for
the best-fitting model of Table~\ref{tab:xillverP} when the reflection
fraction is set  to zero (panel B); when the  depth of the fundamental
cyclotron line is  set to zero (panel  C); when the depths  of the two
harmonics are set to zero (panel D).

During  the fitting  procedure we  noted that  the inner  disk radius,
R$_{in}$, of the reflection  component becomes rather unconstrained by
the  fit because  of  a  strong correlation  with  the other  smearing
spectral parameters  (i.e. inclination  angle and  $\epsilon$). Fixing
the inclination angle to 18$^{\circ}$, 22$^{\circ}$, and 25$^{\circ}$,
we obtained a  lower limit on R$_{in}$ of 90  R$_{g}$, 80 R$_{g}$, and
70 R$_{g}$, respectively.  For higher  values of the inclination angle
R$_{in}$ is not constrained. If  the value of the $\epsilon$ parameter
is  frozen  to 3  (a  value  that  would  correspond to  an  isotropic
point-like irradiating source), we obtained a lower limit for R$_{in}$
of $\sim$\,100 R$_g$; if both the inclination is fixed to 22$^{\circ}$
\citep[upper   limit   based   on  the   calculations   from   ][Table
  3]{rappaport97}  and  $\epsilon$\,=\,3,  then the  lower  limit  for
R$_{in}$ is 120 R$_g$.

In Table~\ref{tab:xillverP}, first column,  we report the best-fitting
results  with  R$_{in}$ fixed  to  a  reference  value of  100  R$_g$.
Finally,  we note  that  if no  systematic error  is  assumed for  the
EPIC/pn data,  then a  strict lower  limit on  R$_{in}$ is  derived at
$\sim$  60  R$_{g}$  (Table \ref{tab:xillverP},  second  column).   We
obtained  similar results  adopting the  EPIC/pn spectrum  without the
three central columns (Table~\ref{tab:xillverP}, third column).

Differently  from  the   \texttt{nthcomp}  Comptonization  model,  the
\texttt{relxill}  continuum has  no low-energy  roll-over, implying  a
higher flux of this component in the softest X-ray band. This affected
the determination of the disk emission, and we obtained a shift of the
\texttt{ezdiskbb} temperature to  lower values and a  higher value for
the disk inner radius (R$_{ez}$ $\sim$ 90 R$_{g}$).  We found that the
reflection  component   requires  a   high  metallicity,   where  iron
over-abundance is $\sim$10  the solar value.  Also  the emission lines
from the  highly ionized transitions  of Si/Ar/Ca were not  fully well
fitted by  the broadband reflection  model and the  residuals strongly
required more  flux at  the line  energies that  we interpreted  as an
indication of possible  higher metal abundance. The  companion star of
GRO  J1744-28 is  an  evolved giant  \citep{gosling07} whose  external
layers are probably rich in the ejecta of the supernova explosion that
generated  GRO  J1744-28  and  a  high level  of  metallicity  in  the
accretion flow could be expected.

The best-fitting parameters  and associated errors are  shown in first
column  of Table~\ref{tab:xillverP},  while  in the  second and  third
columns of the same table,  we comparatively show the two best-fitting
models using  the EPIC/pn  data without  any systematic  error (second
column), and the EPIC/pn spectrum extracted  from the wings of the PSF
(excising  the three  hottest RAWX  columns) and  no systematic  error
(third  column).   The range  of  best-fitting  values and  associated
errors  is  useful  to  comparatively  visualize  the  impact  of  the
assumptions based  on the choice to  assign a systematic error  to the
EPIC/pn data and the impact of possible residual pile-up.

\begin{table*}
\caption{Best-fitting  parameters  and  errors   for  models  using  a
  self-consistent  reflection  component  to fit  the  highly  ionized
  emission  lines.  In  the first  column we  assigned to  the EPIC/pn
  channel  a systematic  error  of  0.5\%, in  the  second column  the
  EPIC/pn spectrum  has no  systematic error, in  the third  column we
  used  the  EPIC/pn  spectrum  obtained by  excising  the  PSF  after
  removing  the  three  central  columns.   Errors  given  at  $\Delta
  \chi^2\,=\,2.706$  (90\%  confidence   for  the  single  interesting
  parameter).}
\begin{tabular}{l lll} 
\hline \hline
  & \texttt{0.5\% sys.}   & \texttt{No sys.error} & \texttt{3 cols} \\
\hline

N$_{\textrm H}$ (10$^{22}$ cm$^{-2}$)  & 6.4$\pm$0.3      & 6.15$\pm$0.15     & 6.2$_{-0.3}^{+0.2}$ \\

$\Gamma$                           & 0.33$_{-0.07}^{+0.12}$   & 0.295$\pm$0.015           & 0.49$\pm$0.07 \\
Norm.$^{a}$                            & 0.45$\pm$0.06    & 0.42$\pm$0.01           & 0.55$_{-0.06}^{+0.04}$ \\
E$_{cut}$  (keV)                       & 8.56$\pm$0.15    & 8.48$\pm$0.10           & 8.94$_{-0.16}^{+0.11}$ \\
\hline
kT$_{disk}$ (KeV)                           &  0.29$\pm$0.04          & 0.30$\pm$0.02  & 0.28$\pm$0.02 \\ 
R$_{ez}$ (R$_g$)   &  90$^{+100}_{-50}$     & 65$_{-15}^{+50}$ & 100$_{-40}^{+35}$ \\
\hline
E$_{1}$ (keV)                         & 4.68$\pm$0.05 & 4.75$_{-0.02}^{+0.03}$ & 4.59$\pm$0.06 \\
W$_{1}$ (keV)                         & 0.68$\pm$0.08 & 1.02$\pm$0.05 & 0.84$\pm$0.07 \\
D$_{1}$ ($\times$ 10$^{-2}$)          & 8.7$\pm$0.2   & 22$\pm$3   & 16$\pm$4 \\
\hline
E$_{2}$ (keV)                        & 10.4$\pm$0.10 & 11.4$\pm$0.08 & 9.43$\pm$0.09 \\
W$_{2}$ (keV)                       & 0.66$\pm$0.10 & 1.0$\pm$0.4 & 0.94$_{-0.08}^{+0.13}$ \\
D$_{2}$ ($\times$ 10$^{-2}$)        & 14$\pm$9      & 24$\pm$15 &  25$\pm$7 \\
\hline
E$_{3}$ (keV)                       & 15.8$_{-0.7}^{+1.3}$   & 16.0$\pm$0.07  &  15.4$\pm$0.7\\
W$_{3}$ (keV)                       & 2.6$_{-0.9}^{+0.7}$    & 2.3$\pm$0.7 &  2.6$\pm$0.8  \\
D$_{3}$ ($\times$ 10$^{-2}$)        & 90$\pm$30              & 93$\pm$22 &  95$\pm$25 \\
\hline
$\epsilon$ (emissivity)             & 1.5$_{-1.5}^{+2.0}$ & 1.1$_{-0.9}^{+1.2}$ & 2.3$_{-0.8}^{+1.2}$ \\
Inclination (deg)                        & 28$_{-10}^{+20}$    & 28$\pm$4            & 23$_{-4}^{+10}$     \\
R$_{in}$ (R$_g$)                    & 100$^{b}$           & $>$60               & 100$^{b}$ \\
\hline
Relxill Log($\xi$)                    & 3.09$\pm$0.02      & 3.09$\pm$0.02    & 3.1$\pm$0.03     \\
Relxill $A_{Fe}$                     & $>$8$^{c}$         & $>$9$^{c}$     & $>$7$^{c}$  \\
Relxill $Refl.~fr.$  (10$^{-2}$)    & 5.3$\pm$0.4      &  5.3$_{-0.2}^{+0.7}$       & 7.6$_{-1.8}^{+1.2}$ \\ 
\hline
\sxvi~ norm.$^{d}$               & 7$\pm$5       & 5.6$\pm$1.9 & 4.9$\pm$3.9  \\
\arxviii~ norm.$^{d}$            & 8$_{-4}^{+2}$ & 7.9$\pm$1.8 & 7.3$\pm$2.5  \\
\caxix~ norm.$^{d}$              & 8$_{-3}^{+2}$ & 7.3$\pm$1.3 & 6.2$\pm$2.2 \\
\caxx~ norm.$^{d}$               & 6$\pm$3       & 5.4$\pm$1.3 & 7.7$\pm$2.0 \\
\hline
C$_1$ PN/RGSo1         & 1.15$\pm$0.02  & 1.16$\pm$0.01 & 1.07$\pm$0.02 \\
C$_2$ PN/RGSo2         & 1.02$\pm$0.03  & 1.10$\pm$0.02 & 1.02$\pm$0.02 \\
C$_3$ PN/JEMX1         & 1.19$\pm$0.02  & 1.18$\pm$0.02 & 1.36$\pm$0.02 \\
C$_4$ PN/JEMX2         & 1.32$\pm$0.02  & 1.32$\pm$0.02 & 1.52$\pm$0.02  \\
C$_5$ PN/ISGRI         & 1.20$\pm$0.03  & 1.16$\pm$0.03 & 1.41$\pm$0.05  \\
\hline
$\chi^2_{red}$ (dof)    & 1.025 (519)  & 1.408 (519) & 1.284 (519) \\
\hline
\hline
\multicolumn{4}{l}{$^a$ In units of photons keV$^{-1}$ cm$^{-2}$ s$^{-1}$ at 1 keV.}\\
\multicolumn{4}{l}{$^b$ Frozen parameter during fitting search.}\\
\multicolumn{4}{l}{$^c$ Hard limit for this parameter set to 10.}\\
\multicolumn{4}{l}{$^d$ Line normalization in units of 10$^{-4}$ photons cm$^{-2}$ s$^{-1}$ in the line.}\\
\end{tabular}
\label{tab:xillverP}
\end{table*}

\subsection{The pulsed fraction in the iron region: comparison with best-fitting models} \label{sect:pulsedfraction}

\begin{figure}
\centering
\includegraphics[height=\columnwidth, width=0.8\columnwidth,  angle=-90]{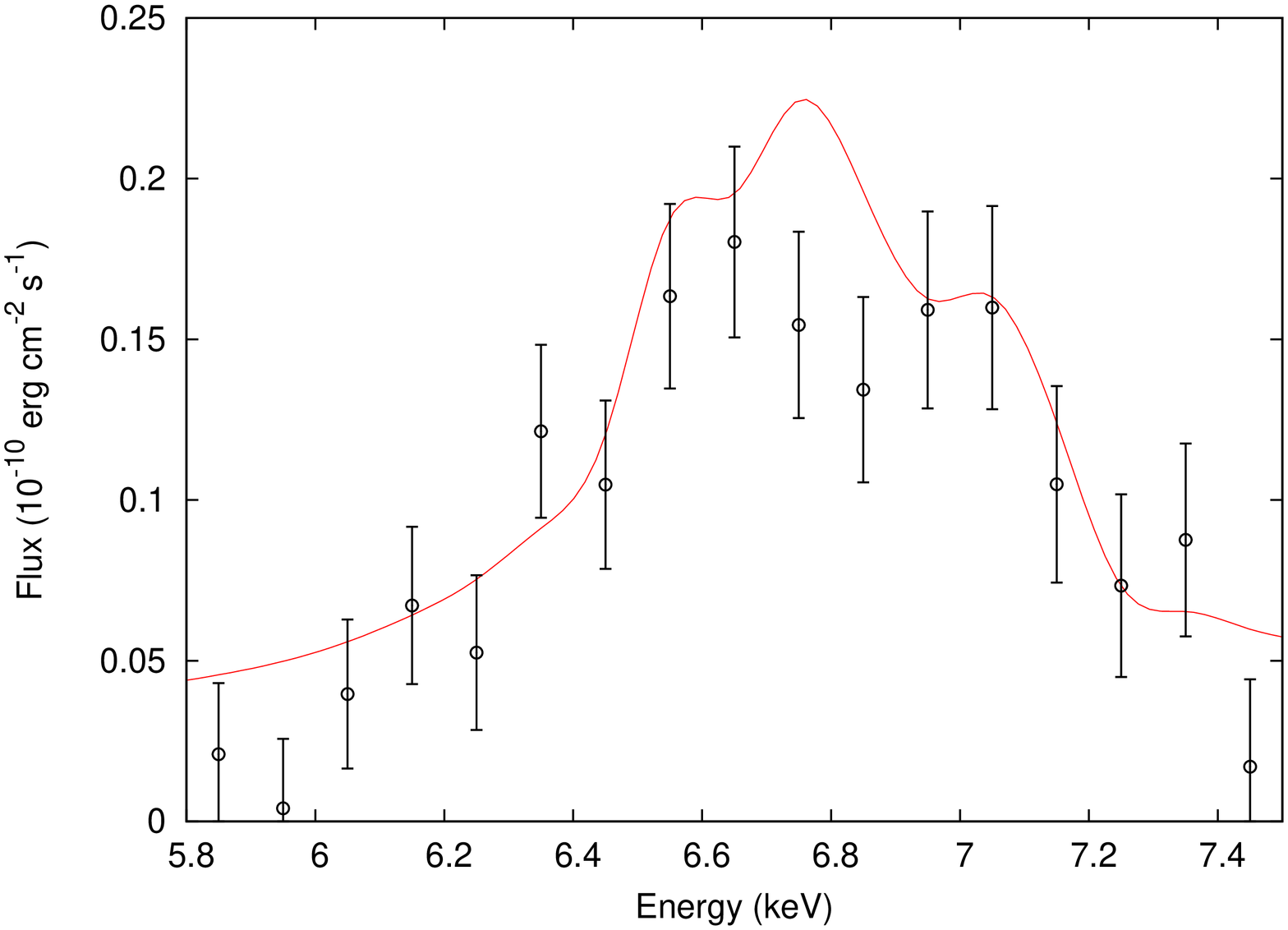}
\caption{Red  line: the  reflection  component flux  according to  the
  best-fit model of Table~\ref{tab:xillverP}.  Black data: excess flux
  from the amplitude fraction according to Eq.~\ref{eq:excessflux}.}
\label{flux_ampl}
\end{figure}

We have shown in Sect.~\ref{timinganalysis} that there is a clear drop
in  the pulsed  fraction  in  the iron  region  of  the spectrum  (see
Fig.~\ref{amplitude}) and  that the  shape of  the drop  is consistent
with  a broad  Gaussian  profile. Such  drop has  been  also noted  in
\citet{nishiuchi99}, who proposed its origin in the not-pulsed flux of
a relativistically  smeared iron line.   We have shown that  the large
iron residuals can be interpreted  as reflection from a truncated disk
and we  establish here if the  shape and strength of  the residuals of
Fig.~\ref{amplitude} are consistent with this origin.  To this aim, we
divided the energy spectral range in bins of 100 eV width from 5.5 keV
to 7.5  keV. We calculate  for each channel  the observed flux  of the
reflection   component,  \texttt{relxill},   and  of   the  reflection
component     according    to     the     best-fitting    model     of
Table.~\ref{tab:xillverP}.  We calculated the not-pulsed $excess$ flux
($F_{ex}$) for the  amplitude versus energy plot  using a best-fitting
third-order polynomial function fit to the pulsed fraction outside the
5.5--7.5  keV   range  as  explained   in  Sect.~\ref{sect:amplitude},
according to the following expression:

\begin{equation} \label{eq:excessflux}
F_{ex} = \left(  \frac{ f_{exp} } { f_{obs} } - 1 \right) \times F_{tot}
\end{equation} 

where $f_{exp}$ is the value of  the $expected$ pulse fraction in that
energy bin from  the best-fitting function, $f_{obs}$  is the observed
pulsed  fraction, and  $F_{tot}$ is  the continuum  flux per  bin.  In
Fig.~\ref{flux_ampl}, we show the results of this procedure.  We found
that the  not-pulsed flux excess  is in reasonable agreement  with the
flux of the reflection component.

\section{Pulse-phase resolved spectroscopy}

\begin{figure}
\centering
\includegraphics[height=0.8\columnwidth, width=\columnwidth]{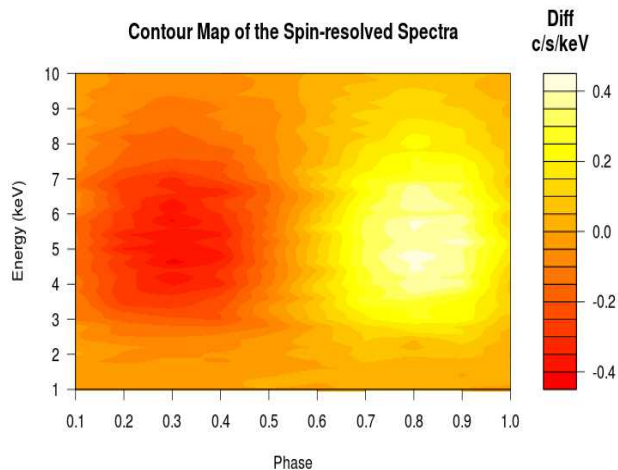}
\caption{Heat map for  the EPIC/pn spin-resolved spectra.   We plot on
  the  y-axis   the  count/s/keV   difference  with  respect   to  the
  time-averaged spectrum.}
\label{heatmap}
\end{figure}

Based on  the best  time-averaged pulse  period during  the persistent
emission,   we   extracted   ten   pulse-resolved   spectra,   equally
phase-spaced, to  study the  spectral evolution as  a function  of the
spin  phase. We  used only  data from  the EPIC/pn  instrument in  the
2.0--9.5  keV  range.  To  get  a  preliminary  idea of  the  spectral
changes,   we  first   plotted  the   count  difference   between  the
time-averaged spectrum  and the phase-selected spectrum  as a function
of the energy  channel.  The result of this operation  is shown in the
heat  map  of  Fig.\ref{heatmap},  where no  evident  substructure  is
visible, implying that most of the spectral changes is due to a slowly
changing slope  of the  overall continuum.   To fit  the spin-resolved
spectra we adopted the  best-fitting model of Table~\ref{tab:xillverP}
and we found  a global acceptable fit (reduced  $\chi^2$ 1287/1241) by
setting  the  following parameters  free  to  vary: the  photon  index
($\Gamma$), the  normalization of the \texttt{relxill}  component, and
the reflection  fraction.  The  remaining parameters  of the  fit were
kept fixed to  the best-fitting values of  the pulse-averaged spectrum
(Table~\ref{tab:xillverP}).

\begin{figure}
\centering
\includegraphics[height=\columnwidth , width=0.8\columnwidth ,  angle=-90]{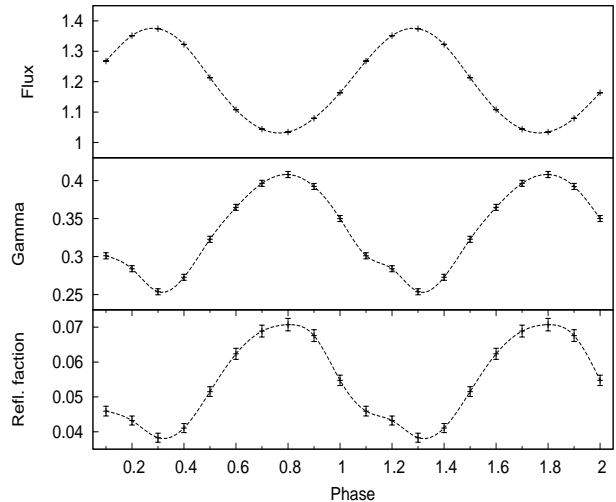}
\caption{Best-fitting  parameters 1 $\sigma$ errors for the  ten phase-resolved
  spectra. From top to bottom: \texttt{relxill} 0.5--10 keV unabsorbed flux in units
  of 10$^{-8}$ erg cm$^{-2}$ s$^{-1}$, photon-index, and reflection fraction.}
\label{fig_pulseresolved}
\end{figure}

In  Fig.~\ref{fig_pulseresolved},   we  show  the  variation   of  the
photon-index, reflection fraction  and absorbed flux as  a function of
the  spin phase,  all following  the  smooth sinusoidal  trend of  the
pulsed  flux.  We  clearly  observe  that  the  lowest  value  of  the
photon-index is found in correspondence  with the pulse maximum, while
the highest reflection  fraction is at the pulse  minimum. Because the
reflection fraction is the ratio between the reflected and the direct,
primary flux,  this result suggests  that the reflected flux  does not
sensibly depend on the pulse phase and we verified this by calculating
the  reflected  flux  for  each phase-selected  spectrum,  finding  it
constant within a $\sim$\,5\% relative  uncertainty. We did not detect
statistically significant variations of the fundamental cyclotron line
parameters along the pulse phase.

\section{Discussion}

In  this paper  we  presented  a complete  analysis  of the  broadband
(1.3--70 keV)  spectral characteristics of the  persistent emission of
the  bursting  pulsar   GRO  J1744-28  based  on   a  long  continuous
\textit{XMM-Newton}  observation  and  on   the  hard  X-ray  coverage
provided by  two of the instruments  on board of $INTEGRAL$  (JEMX and
ISGRI).

The  EPIC/pn  spectrum   is,  however,  affected  by   some  level  of
uncertainty,  because  of  the  high  registered  count  rate  causing
instrumental  spectral  distortions and  shifts  in  the energy  scale
\citep[i.e.    charge  transfer   inefficiency   and  X-ray   loading,
  see][]{pintore14} and  by some level  of photon pile-up.  To process
the EPIC/pn data  we adopted the \texttt{RDPHA}  pipeline as suggested
by the EPIC/pn hardware team, and we obtained our best-fitting results
considering    both   an    excised    PSF   and    the   total    PSF
(\textit{full~spectrum},  see   Table~\ref{tab:xillverP}).   The  main
spectral differences  in the two  spectra affect the  determination of
the power-law photon  index, the cut-off energy, and the  depth of the
first harmonic,  which resulted each other  not compatible.  Moreover,
the  EPIC/pn  spectrum  is  according  to us  affected  by  two  other
(possible)  instrumental   artefacts:  a   narrow  emission   line  at
$\sim$\,2.2 keV, observed  in the spectra of bright  sources either in
emission  or in  absorption \citep{hiemstra11,pintore14},  due to  the
strong effective area derivative of the Au-M edge and a narrow line at
7.11 keV, where there is a  physical spectral discontinuity due to the
sharp  neutral   iron  edge.  The  \texttt{RDPHA}   method  assigns  a
$\sim$\,20 eV  uncertainty in the photon  energy scale reconstruction,
and the appearance of such narrow  feature could be caused by a slight
mismatch in the energy scale reconstruction of photons in this band.

However, despite these caveats, the overall picture and interpretation
of the  main physical processes  producing the X-ray  spectrum remains
quite solid,  because it  was found  not to  sensibly depend  on these
instrumental uncertainties (Table \ref{tab:xillverP}).

\subsection{Continuum emission and cyclotron lines}

The X-ray  continuum spectrum  of an accreting  pulsar is  governed by
different  physical  process  that  are  all  sensible  to  the  local
condition of the accreting plasma and  to the geometry and strength of
the  NS  magnetic field.   Soft  photons  produced by  bremsstrahlung,
synchrotron and thermal, black-body, emission are Compton up-scattered
to higher  energies either  by the energetic,  free-falling, electrons
(bulk-motion Comptonization) or through thermal Comptonization.  X-ray
pulsars  accreting  above  the critical  luminosity  \citep[$L_{crit}$
  $\approx$\,3\,$\times$\,10$^{36}$  erg s$^{-1}$][]{basko76,becker05}
have  a  radiative  shock,  formed  and  sustained  by  the  Eddington
radiation pressure, and  a significant part of  the scattering process
is  due to  thermal Comptonization.   For sources  below the  critical
luminosity, bulk-motion (or  fist-order Fermi energization) dominates.
Sources in this latter category  show steeper spectra ($\Gamma >$\,2),
whereas sources  of the  former class  appear flatter  ($\Gamma <$\,2,
\citet{becker07}).

Because of  the complexity and  non-linearity of all  these processes,
the  pulsar spectra  have  usually been  fitted with  phenomenological
models, that, however,  provided in most cases  a statistical adequate
description of the data, and a  general common frame for comparing the
spectra  of  different  sources  and detect,  even  very  broad  local
absorption/emission  features  \citep{heindl04}.   We found  that  the
broadband spectrum GRO J1744-28, however, cannot be described in terms
of  simple  power-law  models   (whatever  the  prescription  for  the
exponential  cut-off),  or  combinations  of  black-body  and  cut-off
power-laws, that  are more typical for  bright un-magnetized accreting
NSs.  We  observed a  common  trend  of  residuals, extending  in  the
4.0--10.0 keV range,  giving rise to a spectral curvature  that can be
satisfactorily accounted by  a set of broad features  in absorption at
the  edges of  the  iron  K$\alpha$ region,  together  with  a set  of
moderately broadened features in  emission, that can be satisfactorily
described  with a  self-consistent  reflection  component, smeared  by
dynamical and  relativistic effects  produced at the  inner edge  of a
truncated  disk.   The presence  of  these  features is  statistically
required independently  from the adopted continuum.   We have modelled
it first with  the sum of a  soft thermal disk emission  and a thermal
Comptonization model (\texttt{nthcomp}), that  has two main parameters
that  determine  the  cut-off, high-energy,  roll-over  (the  electron
temperature  and the  cloud  optical depth,  related  to the  $\Gamma$
parameter).  We  note here that  we obtained very similar  results, in
terms   of   $\chi^2$   and   spectral  parameters   for   the   local
emission/absorption    features,   using    a    power-law   with    a
\texttt{highecut} cut-off,  a model widely adopted  for describing the
hard spectra  of X-ray  pulsars.  The addition  of an  other continuum
component (as a black-body  emission) resulted in thermal temperatures
that  closely   followed  the  shape  of   the  \texttt{nthcomp}  soft
seed-photon temperature,  and we  concluded that  the \texttt{nthcomp}
encloses all the  characteristics of the most commonly  models used in
the literature to fit the spectra of bright X-ray pulsars.

In Sect.~\ref{sect:xillver} we showed results from a simpler continuum
made  with  a cut-off  power-law  (\texttt{relxill}  model), a  softer
thermal disk emission (\texttt{ezdiskbb}),  and a broadband reflection
component.   The reflected  spectrum  is also  an important  continuum
contributor, through  continuum Thomson scattering  and bremsstrahlung
emission, and  significant changes  in the best-fitting  parameters of
the  other  local  features   in  absorption  were  expected  (compare
Table~\ref{tab:spectral_fits}  and  Table~\ref{tab:xillverP}).  It  is
clear that the modelling of the underneath continuum is more important
than the  statistical uncertainties  from the  best-fitting modelling,
although the general  picture of the cyclotron  features appears quite
solid. The uncertainty  on the cyclotron fundamental energy  is of the
order  of 6\%,  the  width of  25\%,  whereas the  depth  is the  most
sensible parameter,  due to  the strong  correlation with  the general
shape of the fluorescence iron line.

If we  use the  best-fitting parameters  from Table~\ref{tab:xillverP}
for the  fundamental cyclotron  line energy (E$_{1}$\,=\,4.7  keV), we
can derive the magnetic field  of the NS using Eq.~\ref{eq:cyclotron},
leading       to      a       magnetic      dipole       field      of
(5.27\,$\pm$\,0.06)\,$\times$\,10$^{11}$  G  (magnetic  dipole  moment
$\mu$\,=\,5.27\,$\times$\,10$^{29}$  G cm$^3$),  assuming the  CRSF is
formed at the surface of the NS (gravitational red-shift $z$\,=\,0.3).

This estimate  allows to constrain the  magnetospheric radius ($R_M$),
where  the ram  pressure  of the  accretion flow  is  balanced by  the
magnetic pressure, according to the formula \citep{frankkingraine}:

\begin{equation} \label{eq:magnetorasius}
R_M = 2.9 \times 10^8  \kappa_m  M_{NS}^{1/7} R_6^{-2/7} L_{37}^{-2/7} \mu_{30}^{4/7} \quad \textrm{cm}
\end{equation}

where  $\kappa_m$  is a  coefficient  related  to the  accretion  flow
geometry (1 for  isotropic radial accretion), $M_{NS}$ is  the NS mass
in units of  $M_{\odot}$, $R_6$ is the NS radius  in units of 10$^{6}$
cm,  $L_{37}$ is  the  source  luminosity in  units  of 10$^{37}$  erg
s$^{-1}$, and  $\mu_{30}$ is  the magnetic dipole  moment in  units of
10$^{30}$ G cm$^{3}$.  From our  B estimate, the bolometric luminosity
derived  by  extrapolating  the   observed  flux,  assuming  isotropic
emission,  and  classical values  for  the  NS  mass and  radius  (1.4
M$_{\odot}$,   10$^6$    cm   radius),   we   derive    a   value   of
$R_M$\,$\simeq$\,10$^8$\,$\times \kappa_m$ cm.  The co-rotation radius
for a NS of period P$_{spin}$ is

\begin{equation}
R_{\Omega} = \left( \frac{G M P^2_{spin}}{4 \pi^2} \right)^{1/3}
\end{equation}

and corresponds for GRO J1744-28 to 10$^{8}$ cm. A value of $\kappa_m$
close to 1 (isotropic accretion) would  imply that the NS is accreting
at the  spin equilibrium.  However,  we derived from  the best-fitting
parameters of  the disk  emission and of  the reflection  component an
inner   disk   radius    lower   limit   closer   to    a   value   of
2\,$\times$\,10$^{7}$ cm,  that sets the $\kappa_m$  lower limit value
at  $\sim$  0.2, that  is  in  agreement  with the  value  $\sim$\,0.2
reported  for  low  B-field  pulsars  in  \citet{burderi02},  and  not
unrealistically  lower than  a more  recent estimate  of 0.5  found by
\citet{long05}. We  also note  that, as discussed  by \citet{bozzo09},
the  expected   value  of   this  parameter   is  affected   by  large
uncertainties due  to the different theoretical  prescriptions for the
magnetospheric radius published so far.

We found that the width of  the fundamental CSRF is 15--20\% the value
of the  energy position,  and this  is in line  with what  observed in
similar     other     accreting      pulsars     \citep[see     Fig.~5
  in][]{heindl04}. Adopting the classical  formulation for the thermal
broadening   in   a   plasma    with   electron   temperature   $kT_e$
\citep{meszaros85}:

\begin{equation} \label{eq:meszaros}
\left( \frac{\Delta E}{E} \right)_{FWHM} = \left( \frac{8 \textrm{ln} 2 \quad k T_e}{m_e c^2}   \right)^{1/2} \textrm{cos} \theta 
\end{equation}

assuming  an angle  $\theta$  $\sim$\,20$^{\circ}$, we  can derive  an
electron temperature  of $\sim$\,11 keV,  which is quite close  to the
cut-off  energy of  the broadband  spectrum, estimated  at $\sim$\,8.6
keV.  From the  spin-resolved analysis, we found  that the fundamental
cyclotron feature did not show any statistical significant correlation
with   the  pulsed   emission.    \citet{degenaar14}   noted  in   the
$Chandra$/HETGS observation of GRO J1744-28 an unidentified absorption
line  at  $\sim$\,5.11 keV  that  was  tentatively associated  with  a
cyclotron  absorption feature,  though  the line  width  could not  be
constrained and it  was kept fixed to  10 eV in the  fit.  Because the
energy and the  width of the fundamental cyclotron line  as derived by
our  analysis of  the EPIC/pn  data (Table~\ref{tab:xillverP})  points
towards  a significantly  broader feature,  we retain  that these  two
detections are not related.

We found a  suggestive, but statistically weak, hint  of two cyclotron
harmonics  at  $\sim$\,11  keV  and  $\sim$\,15  keV.   We  note  that
\citet{younes15} reported also a broad dip in the $NuSTAR$ spectrum of
this source at $\sim$\,10 keV, whose  shape resembles the one shown in
the $INTEGRAL$/JEMX data. The widths  and depths of these two features
could not be independently determined,  because the features are broad
and they spectrally overlap.  Moreover,  we noted a certain dependence
based on  the assumptions  for the continuum  and for  the systematics
that affect  the inter-calibration  of the different  instruments that
were   used  to   constrain  these   shapes.   Notwithstanding   these
limitations,  the energies  of  the first  and  second harmonics  were
always found close to the expected  harmonic ratios 1:2 and 1:3 (for a
fundamental line at $\sim$\,4.7 keV), and GRO J1744-28 would be, after
4U  0115+64  \citep[][CRSF  fundamental  at 14  keV]{muller13}  and  V
0332+53 \citep[][CRSF fundamental at 26 keV]{pottschmidt05}, the third
X-ray pulsar  to show a second  harmonic in its spectrum.   The slight
discrepancy in  the harmonic  ratios (2.2  and 3.3  for the  first and
second harmonic, respectively)  has been observed also  in other X-ray
pulsars, and it  has been shown to be dependant  on the accretion rate
as discussed in  \citet{nakajima10}.  We found that  the two harmonics
have larger  depths with  respect to the  fundamental, as  reported in
other   sources   \citep[e.g.   for   X0115+63,][]{santangelo99},   an
observational   fact   explained   through  the   two-photon   process
\citep{alexander91}. Because  these features  lie outside  the EPIC/pn
band,  we could  not track  parameter variations  with respect  to the
spin-phase.

\subsection{The highly ionized reflection component: GRO J1744-28 between the LMXB and the HMXB class}

The set of highly ionized  feature is satisfactorily compatible with a
reflection spectrum from  a cold disk truncate at  large distance from
the NS.  The inner disk radius  estimate is correlated with the values
of the  other smearing  parameters, but for  realistic values  of both
inclination angle  ($\sim$\,22 deg)  and index  of the  emissivity law
profile  ($\epsilon$\,=\,3),  a  lower  limit  value  on  R$_{in}$  is
obtained at 120 R$_g$.  The temperature and the luminosity of the disk
continuum  component   was  found   compatible  with   this  estimate.
Differently  from  \citet{nishiuchi99}, we  do  not  find evidence  of
\textit{extreme} red wings, that would imply low inner disk radii, and
our estimate is  rather close to an independent estimate  based on the
iron  line shape  observed in  the $Chandra$  observation analysed  by
\citet{degenaar14}.

The presence  of reflection components in  highly magnetized accreting
pulsars has been  usually neglected, because of the  large distance of
the  inner  disk  radius  (if present)  and  expected  low  reflection
fraction. \citet{ballantyne12} first proposed a set of self-consistent
reflection tables  and showed that moderately  broad emission features
could be signatures of such component  as in the case of the accreting
pulsar LMC X-4.  In that case however blurring parameters derived from
best-fitting  did not  provide physically  consistent results,  as the
derived  inner radii  were too  low  compared with  the expected  disk
truncation radius at  the magnetosphere.  On the  contrary, for weakly
magnetized  accreting  X-ray  pulsars   the  detection  of  reflection
components has  been already established  in a considerable  amount of
cases. A  relativistic iron line  profile has been first  reported for
SAX J1808.4-3658,  where the  blurring parameters  were found  in line
with  the  expectations  \citep{papitto09},  and later  also  for  IGR
J17511-3057 \citep{papitto10}, IGR J17480-2446 \citep{miller11} and in
HETE J1900.1-2455 \citep{papitto13}.  In the above-mentioned cases the
inner disk radii  were constrained in the range 20--40  R$_g$ and were
consistent with  the possible  magneto-spheric radii derived  from the
study of spin timing, whereas in non-pulsating accreting neutron stars
significantly   lower  values   were  derived   \citep[6--15  R$_{g}$,
  see][]{cackett10}.

We derived an inclination angle between 18$^{\circ}$ and 48$^{\circ}$,
that is  within the range  of values for  the system according  to the
possible  evolutionary histories  of  the system  \citep{rappaport97}.
From the X-ray mass function of GRO J1744-28 \citep{finger96}, we thus
derive an upper  limit on the companion mass  between 0.23 M$_{\odot}$
and  0.27  M$_{\odot}$,  for  a  1.4  and  1.8  M$_{\odot}$  NS  mass,
respectively.  This  range encompasses  the most probable  value (0.24
M$_{\odot}$)   indicated   for   the   present   companion   mass   by
\citet{rappaport97}.

The reflected component has a  high ionization degree (log$\xi$ $\sim$
3.1) as expected  from the presence of H- and  He-like emission lines,
and from the  distance of the inner disk radius  and bolometric source
luminosity,   we  can   estimate   the   reflector  electron   density
$n_{e}$\,=\,$L_{bol}   /  (r^2   \xi)$  $\sim$\,3\,$\times$\,10$^{20}$
cm$^{-3}$.

\section{Conclusions}

We  presented results  from  the spectral  analysis  of the  broadband
spectrum    of   the    bursting   pulsar    GRO   J1744-28.     Using
$XMM-Newton/INTEGRAL$ data  we covered  the 1.3--70 keV  energy range,
when  the source  was accreting  near the  Eddington limit  for a  1.4
M$_{\odot}$ NS.  Independently from the adopted continuum, we found in
the spectrum  a moderately broadened  absorption feature in  the range
4.6--5.3 keV and a set  of moderately broadened emission features.  We
interpreted  this absorption  dip  in terms  of  a cyclotron  resonant
scattering feature and we also  found suggestive evidence of two other
absorption features close  to the energies expected for  the first and
second harmonic. From this interpretation, we derived an estimation of
the  GRO J1744-28  pulsar magnetic  field, 5.27\,$\pm$\,0.06\,$\times$
10$^{11}$ G.\\  We located the  region responsible for  the moderately
broadened  emission  lines  outside  the pulsar  magnetosphere,  in  a
reflecting, accretion disk.  The study of the pulse amplitude fraction
and  the spin-resolved  spectra  indicated that  the  iron complex  in
emission is not pulsed and its  flux is not phase-dependent.  We found
that  the  disk  surface  is  highly  photo-ionized,  with  log($\xi$)
$\sim$\,3.1, with  a metal over-abundance  of a factor  $\sim$\,10 and
truncated  at  a distance  of  $\gtrsim$\,100  R$_g$, where  dynamical
effects, caused  by the  high velocities  of matter  in the  disk, are
still  important. The  spectrum  is also  characterized  by soft  disk
emission, whose truncation radius results compatible with the estimate
from the  reflected component. These two  independent, but converging,
constraints allowed  us to estimate the  Alfv\'en efficiency parameter
for disk-accretion, close  to 0.2. The lower limit  on the inclination
angle  of the  disk is  18$^{\circ}$ and  compatible with  independent
estimates  on   the  possible  evolutionary  history   of  the  system
\citep{rappaport97}.

\section{Acknowledgements} 
\small We  are specially  grateful to N.  Schartel, who  made possible
this ToO observation  through the Director Discretionary  Time and all
the  \textit{XMM-Newton}   team  who  performed  and   supported  this
observation.\\   The  High-Energy   Astrophysics   Group  of   Palermo
acknowledges  support from  the Fondo  Finalizzato alla  Ricerca (FFR)
2012/13,  project  N.  2012-ATE-0390,  founded  by  the University  of
Palermo.\\  A.\  R.\  gratefully acknowledges  the  Sardinia  Regional
Government  for the  financial support  (P.\ O.\  R.\ Sardegna  F.S.E.
Operational Programme  of the Autonomous Region  of Sardinia, European
Social Fund 2007-2013  - Axis IV Human Resources,  Objective l.3, Line
of Activity l.3.1).

\bibliographystyle{mn2e}
\bibliography{refs}

\end{document}